# Entanglement parallelization via quantum Fourier transform

**Mario Mastriani**

Knight Foundation School of Computing & Information Sciences, Florida International University,
11200 S.W. 8th Street, Miami, FL 33199, USA

ORCID Id: 0000-0002-5627-3935
E-mail: mmastria@fiu.edu

**Summary** In this study, we present a technique based on the quantum Fourier transform (QFT) that allows the generation of disjoint sets of entangled particles, in such a way that particles of the same set are entangled with each other, while particles of different sets are completely independent. Several applications of this technique are implemented on three physical platforms, of 5 (Belem), 7 (Oslo), and 14 (Melbourne) qubits, of the international business machine (IBM Q) quantum experience program, where all these applications were specially selected due to their particular commitment to the future Quantum Internet.

## 1 INTRODUCTION

In recent years, the close link between quantum entanglement [1-3] and the quantum Fourier transform (QFT) [4, 5] has been unequivocally established [6-10]. From these studies, not only the recognition of entanglement arises as an eminently spectral phenomenon, but also the notion of the quantum Fourier transform (QFT) as the natural binding block to create entanglement is incorporated in the specialty literature [8], given that until that moment, the greatest contribution of QFT had to do with its applications in phase estimation [11], in particular, applied to the Shor algorithm to factor in prime numbers [12]. In these works [6-10], it is also shown that all the quantum communications protocols [13] to be used in the future Quantum Internet [14-23] have the QFT as structural support. In this sense, we can mention the quantum teleportation protocol [24-29], as well as all the protocols involved in quantum cryptography [30], among which: quantum secret sharing (QSS) [31], quantum key distribution (QKD) [32-38] (with and without entangled photons, which require quantum repeaters [39, 40] in both their terrestrial and space implementations to extend their range), and quantum secure direct communications (QSDC) [41] (which currently can operate with or without entangled photons and does not require the use of keys to encrypt the message to be transmitted) stand out. However, we must also take into account the presence of QFT in all quantum gates [6-10] used in quantum computing [5, 42].

The central idea of this study is to understand the most essential aspects of the aforementioned protocols based on a simultaneous analysis that involves the temporal domains (which constitutes the traditional approach present in the literature) and the spectral domain based on QFT [6-10], to improve the performance of those protocols, or better yet, develop new protocols that are more resistant to all types of operational errors, i.e., bit-flip, phase-flip and phase-change error that can occur in a noisy channel.

On the other hand, several implementations of the quantum communications and cryptography protocols involving the intervention of the QFT [6-10], showed similar performance both on physical platforms with free access via the web such as

IBM Q Experience [43], Rigetti Computing [44], and Quantum Inspire by QuTech [45], as well as on simulators like Quirk [46], Quantum Programming Studio [47], and Qiskit [43]. Therefore, for the present study, it is appropriate to consider the implementations of the aforementioned protocols on physical machines of 5 (Belem), 7 (Oslo), and 14 (Melbourne) qubits of the IBM Q Experience program [43] are sufficiently conspicuous by themselves, without the need to resort to complicated disambiguating analyses of the multi-platform type.

As it will be seen in this study, the possibility of working with an emitting source of disjoint sets of entangled particles [10] will free us from resorting to complicated configurations of the Greenberger–Horne–Zeilinger (GHZ) type [5] for large numbers of entangled photons at the same time, which are so difficult to implement in practice. In the context of the Quantum Internet [14-23], this possibility will constitute an invaluable tool to implement ring-type urban or satellite topologies, which will be particularly useful when carrying out key relays for QKD [32-38].

Other applications of entanglement parallelization are a) multi-orbit architecture [48] for early intrusion detection and space quantum ring surveillance, b) quantum cooperative multicast in a quantum hybrid topology network [49] for communications in large urban areas, and c) as a replacement for GHZ states in multicast QKD, because it eliminates inter-channeling, as well as coupling noise among channels.

The outline of the paper is as follows: In Section 2, the theoretical foundation of entanglement parallelization via quantum Fourier transform for sets of different numbers of entangled particles is introduced. Section 3 presents four families of applications with a fundamental projection on the future Quantum Internet. Finally, Section 4 deals with the general conclusions of this study.

## 2 THEORETICAL FOUNDATION

In this section, we will analyze various configurations of parallel entanglement technology, which will be presented in ascending order of complexity, where the generic case corresponds to a single source of $M$ independent sets of $N$ entangled particles in each set. This can be symbolically expressed as follows:

$$S_N^M \rightarrow \left(p_0^0, p_1^0, \ldots, p_{N-1}^0\right) \cup \left(p_0^1, p_1^1, \ldots, p_{N-1}^1\right) \cup \ldots \cup \left(p_0^{M-1}, p_1^{M-1}, \ldots, p_{N-1}^{M-1}\right) \\ = \bigcup_{m=0}^{M-1} \left(p_0^m, p_1^m, \ldots, p_{N-1}^m\right), \quad (1)$$

where $S_N^M$ represents the source of $M$ sets of $N$ entangled particles per set, $p_n^m$ is the $n^{\text{th}}$ entangled particle of the $m^{\text{th}}$ set, where $n \in [0, N-1]$, and $m \in [0, M-1]$, and $\cup$ is the operator known as *the union of sets*. Then, $\bigcup_{m=0}^{M-1}$ represents the union of the $M$ sets of the source.

An equivalent way to express Eq. (1) is through the following equation,

$$S_N^M \rightarrow \frac{\left(\left|0_0^0 0_1^0 \ldots 0_{N-1}^0\right\rangle + \left|1_0^0 1_1^0 \ldots 1_{N-1}^0\right\rangle\right)}{\sqrt{2}} \cup \frac{\left(\left|0_0^1 0_1^1 \ldots 0_{N-1}^1\right\rangle + \left|1_0^1 1_1^1 \ldots 1_{N-1}^1\right\rangle\right)}{\sqrt{2}} \cup \ldots \cup \frac{\left(\left|0_0^{M-1} 0_1^{M-1} \ldots 0_{N-1}^{M-1}\right\rangle + \left|1_0^{M-1} 1_1^{M-1} \ldots 1_{N-1}^{M-1}\right\rangle\right)}{\sqrt{2}} \\ = \bigcup_{m=0}^{M-1} \frac{\left(\left|0_0^m 0_1^m \ldots 0_{N-1}^m\right\rangle + \left|1_0^m 1_1^m \ldots 1_{N-1}^m\right\rangle\right)}{\sqrt{2}}, \quad (2)$$

where $|0\rangle = [1 \ 0]^T$ is the *spin-up* (or North pole on the Bloch sphere), $|1\rangle = [0 \ 1]^T$ is the *spin-down* (or South pole on the Bloch sphere), $(\bullet)^T$ means *transposed of* $(\bullet)$, while $|0\rangle_n^m$ is the $n^{\text{th}}$ entangled *spin-up* of the $m$ set and $|1\rangle_n^m$ is the $n^{\text{th}}$ entangled *spin-down* of the $m$ set.

The generic form of Eq. (2) turns out to be,

$$S_N^M \rightarrow \bigcup_{m=0}^{M-1} \frac{\left(|0_m\rangle^{\otimes N} + |1_m\rangle^{\otimes N}\right)}{\sqrt{2}}. \quad (3)$$

Finally, Fig. 1 represents what is expressed in Eqs. (2), and (3), where particles of the same set are mutually entangled, while we can graphically associate the mutual independence among particles of the different $M$ sets with the total absence of intersection between those sets.

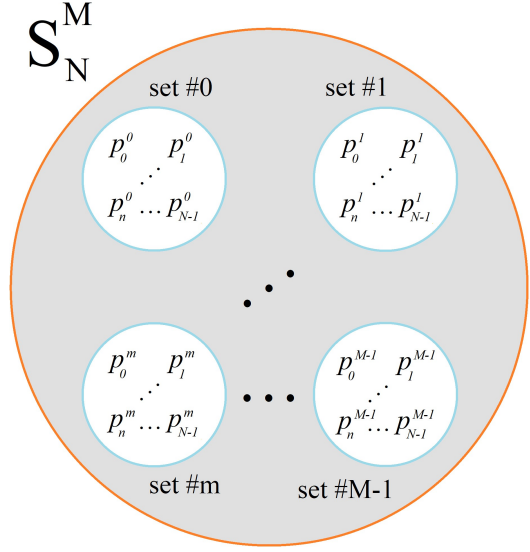

**FIGURE 1** Source $S_N^M$ of *M* independent sets of *N* entangled particles each. The figure only shows four sets (so as not to complicate the drawing) with no intersection areas between them, i.e., they are disjoint sets, which means that the correlation between particles from different sets is null.

## 2.1 Generation of disjoint sets of entangled pairs

The generation of disjoint entangled pairs, i.e., pairs completely independent of other pairs from the same source *S*, can be seen in the examples in Fig. 2, where figures (a, and b) correspond to 2, (c) to 4, and (d) to *N* entangled pairs, completely independent of each other. Specifically, in Fig. 2(a), qubits q[0] and q[2] are entangled, as qubits q[1] and q[3] are, while the pair {q[0], q[2]}, is completely independent of the pair {q[1], q[3]}. In Fig. 2(b), {q[0], q[1]} are entangled, the same as {q[2], q[3]}, although both pairs are completely independent of each other. Finally, Fig. 2(c) is a particular case of Fig. 2(d), and in the latter, only the pairs {q[$i$], q[N+$i$]} / $i \in$ [0, N-1] are entangled.

Figures 2(a) and (b) are completely equivalent. However, the second will greatly facilitate the theoretical deduction, which we will only carry out in this study for this simple case due to the inherent equational complexity as the number of entangled pairs increases. Without losing generality, all the configurations in Fig. 2 generate pairs of the same Bell state [1-3, 5],

$$|\beta_{00}\rangle = |\Phi^+\rangle = 1/\sqrt{2}(|00\rangle + |11\rangle) = CNOT(H \otimes I_{2\times2})|00\rangle, \qquad (4)$$

where

$$CNOT = \begin{bmatrix} 1 & 0 & 0 & 0 \\ 0 & 1 & 0 & 0 \\ 0 & 0 & 0 & 1 \\ 0 & 0 & 1 & 0 \end{bmatrix} \qquad (5)$$

is the two-qubit gate known as *Controlled-NOT*, or *CNOT* [5], and $I_{2\times2}$ is the identity matrix [5],

$$I_{2\times2} = \begin{bmatrix} 1 & 0 \\ 0 & 1 \end{bmatrix}, \qquad (6)$$

while *H* is the Hadamard gate [5],

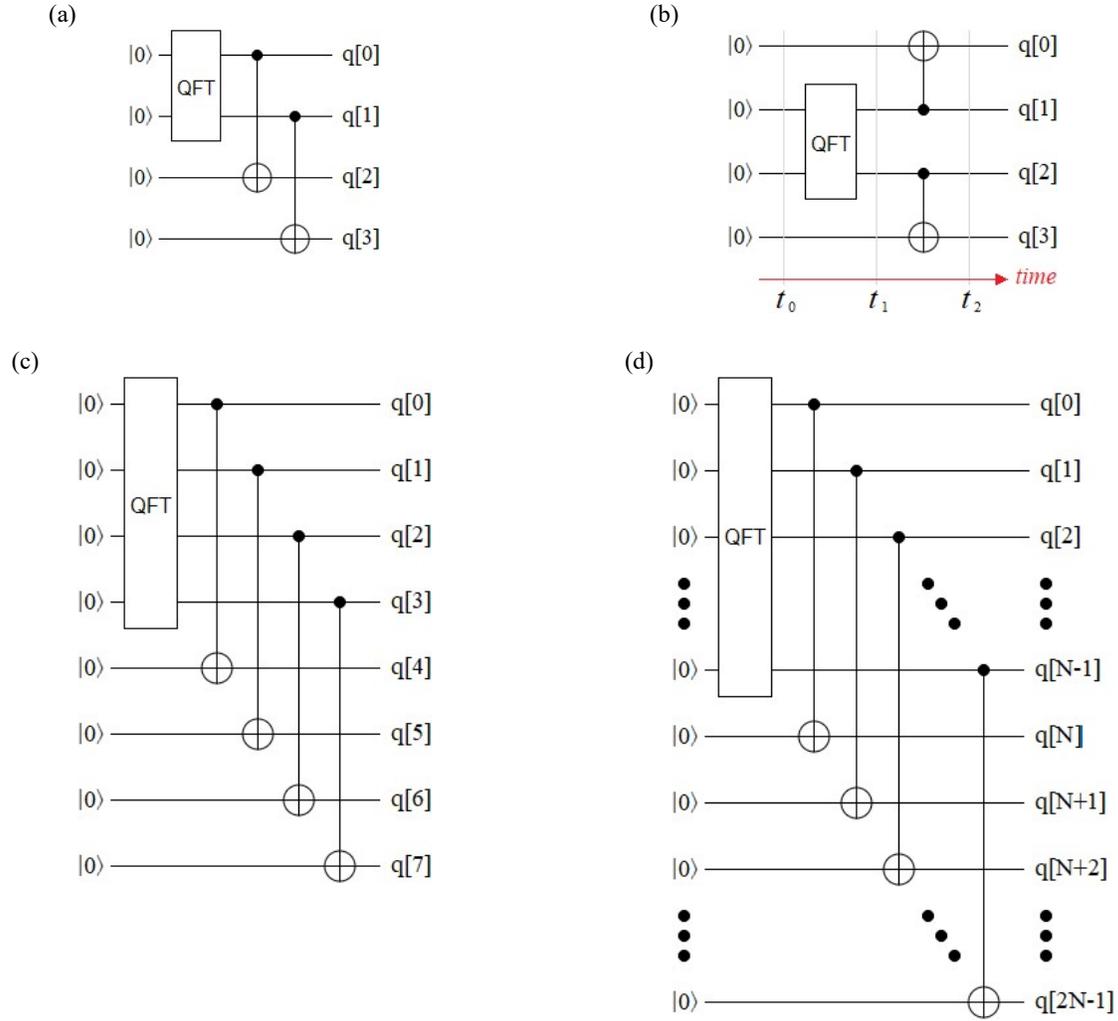

**FIGURE 2** Three examples of single sources of entangled pairs $|\beta_{00}\rangle$ independent of each other: a) $S_2^2$ of two entangled pairs $|\beta_{00}\rangle$ independent of each other, b) Same as above but using a $CNOT_{flipped}$ gate, c) $S_2^4$ of four entangled pairs $|\beta_{00}\rangle$ independent of each other, and d) $S_2^N$ of $N$ entangled pairs $|\beta_{00}\rangle$ independent of each other.

$$H = \frac{1}{\sqrt{2}}\begin{bmatrix} 1 & 1 \\ 1 & -1 \end{bmatrix}. \tag{7}$$

Moreover, Fig. 2(b) is completed with a 2-qubit QFT block, and a *flipped-CNOT* gate, or $CNOT_{flipped}$, which should not be confused with its reverse $CNOT^{-1} = CNOT$, where,

$$QFT = \frac{1}{2}\begin{bmatrix} 1 & 1 & 1 & 1 \\ 1 & i & -1 & -i \\ 1 & -1 & 1 & -1 \\ 1 & -i & -1 & i \end{bmatrix}, \text{ where } i = \sqrt{-1}, \tag{8}$$

and

$$CNOT_{flipped} = \begin{bmatrix} 1 & 0 & 0 & 0 \\ 0 & 0 & 0 & 1 \\ 0 & 0 & 1 & 0 \\ 0 & 1 & 0 & 0 \end{bmatrix}. \tag{9}$$

From here on, we will carry out the theoretical deduction based on the red timeline of Fig. 2(b), with:

$$|\psi(t_0)\rangle = |0000\rangle \tag{10}$$
$$= [1\ 0\ 0\ 0\ 0\ 0\ 0\ 0\ 0\ 0\ 0\ 0\ 0\ 0\ 0\ 0]^T,$$

$$|\psi(t_1)\rangle = (I_{2\times 2} \otimes QFT \otimes I_{2\times 2}) \times |\psi(t_0)\rangle \tag{11}$$
$$= [1/2\ 0\ 1/2\ 0\ 1/2\ 0\ 1/2\ 0\ 0\ 0\ 0\ 0\ 0\ 0\ 0\ 0]^T,$$

where "×" is the matrix product, and "⊗" is the Kronecker product [5]. Finally,

$$|\psi(t_2)\rangle = (CNOT_{flipped} \otimes CNOT) \times |\psi(t_1)\rangle \tag{12}$$
$$= [1/2\ 0\ 0\ 1/2\ 0\ 0\ 0\ 0\ 0\ 0\ 0\ 0\ 1/2\ 0\ 0\ 1/2]^T$$
$$= |\beta_{00}\rangle \otimes |\beta_{00}\rangle.$$

Equation (12) verifies that the configuration in Fig. 2(b) constitutes a single source $S_2^2$ delivering two independent pairs of entangled particles, where the qubits q[0] and q[1] share a Bell state $|\beta_{00}\rangle$ [5], and the qubits q[2] and q[3] share another one. Figure 2 (c) represents a single source $S_2^4$ of four independent pairs of entangled photons, with Bell states of type $|\beta_{00}\rangle$ between the qubits q[0] and q[4], q[1] and q[5], q[2] and q[6], and q[3] and q[7]. Finally, Fig. 2(d) shows a single source $S_2^N$ of N independent pairs of Bell states of type $|\beta_{00}\rangle$.

## 2.2 Generation of disjoint sets of GHZ$_N$ states

In this family of application examples of entanglement parallelization, we can consider the Bell state $|\beta_{00}\rangle$ of Eq. (4) as a particular case of the family of states $|GHZ_N\rangle$ [1-3, 5], specifically, $|\beta_{00}\rangle = |GHZ_2\rangle$, i.e, a case of $|GHZ_N\rangle$ states of only two entangled particles. In this way and in order to explore the ductility of the technique presented here, we will compare from lowest to highest different cases of single sources of independent sets of entangled particles in a complementary way to the examples developed in the previous subsection. Therefore, we turn directly to Fig. 3, where in all cases we are dealing with single sources of two independent sets of entangled particles of type $|GHZ_N\rangle$,

$$|GHZ_N\rangle = \frac{\left(|0\rangle^{\otimes N} + |1\rangle^{\otimes N}\right)}{\sqrt{2}}, \tag{13}$$

in such a way that Eq. (3) can be rewritten as,

$$S_N^M \to \bigcup_{m=0}^{M-1} |GHZ_N\rangle^m. \tag{14}$$

Equation (14) represents a source of M disjoint sets of $|GHZ_N\rangle$ states [5]. These types of states are of vital importance in the implementation of modern cryptographic protocols [51]. The contribution of the parallelization of these types of states from a single source gives rise to the creation of more efficient Network Operations Center/Command (NOC), to connect multiple satellites at the same time [52]. At the end of this paper, we will see an example of this.

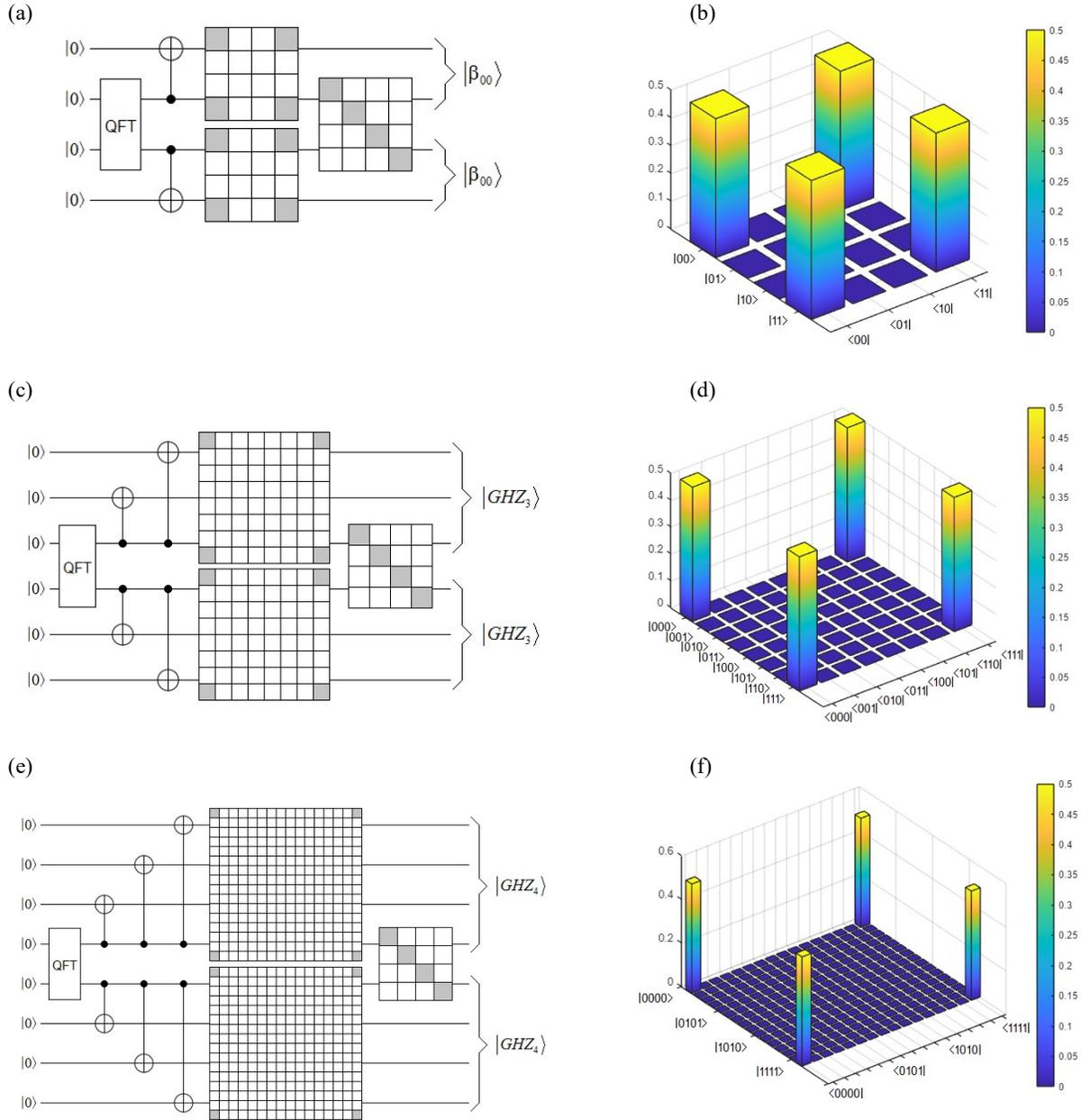

**FIGURE 3** Three examples of two independent sets of different numbers of entangled particles each: a) represent the configurations of Figs. 2(a) and (b), i.e., for a source $S_2^2$, where the two upper and two lower qubits have their respective density matrices (grey) indicating the presence of entanglement, however, the two central qubits have a density matrix of the diagonal type which indicates that both $|\beta_{00}\rangle$ pairs are mutually independent, b) density matrix in 3D bars between the two upper qubits, as well as between the two lower ones in case (a), c) Idem in case (a) but for a source $S_3^2$ that generates two independent $|GHZ_3\rangle$ triads of entangled particles, d) density matrix in 3D bars between the three upper qubits, as well as between the three lower ones in case (c), e) Idem to case (c) but for a source $S_4^2$ that generates two independent sets $|GHZ_4\rangle$ of 4 entangled particles each, f) density matrix in 3D bars between the four upper qubits, as well as between the four lower ones in case (e).

Figure 3(a) shows a single source $S_2^2$, the same as the one represented in Fig. 2(b). It can be seen that both the two upper and the two lower qubits have density matrices that unequivocally represent states of the $|\beta_{00}\rangle$-type, i.e., entangled states, although the density matrix involving the two central qubits, of the diagonal type, gives us indication that both pairs of qubits are completely independent. In the three 2D density matrices, a gray element represents a non-zero value, while a white element represents a zero. Figure 2(b) gives specific values to the elements of the 3D density matrices of the Bell state $|\beta_{00}\rangle$. With the same criteria as in the case of Fig. 3(a), Fig. 3(c) presents a single source $S_3^2$ of two sets of three entangled states of type $|GHZ_3\rangle$ each, where both sets are completely independent of each other. Figure 3(d) shows the density matrix of states $|GHZ_3\rangle$ in 3D bars. Something similar happens with Fig. 3(e) for the case of a single source $S_4^2$ of two independent sets of four entangled states of type $|GHZ_4\rangle$ each, where Fig. 3(f) represents the density matrix of states $|GHZ_4\rangle$ in 3D bars.

The possibility of implementing single sources $S_N^M$ for the generation of $M$ parallel $|GHZ_N\rangle$ states gives rise to different configurations of secure communications based on the quantum secret sharing (QSS) protocol [31], which will be addressed

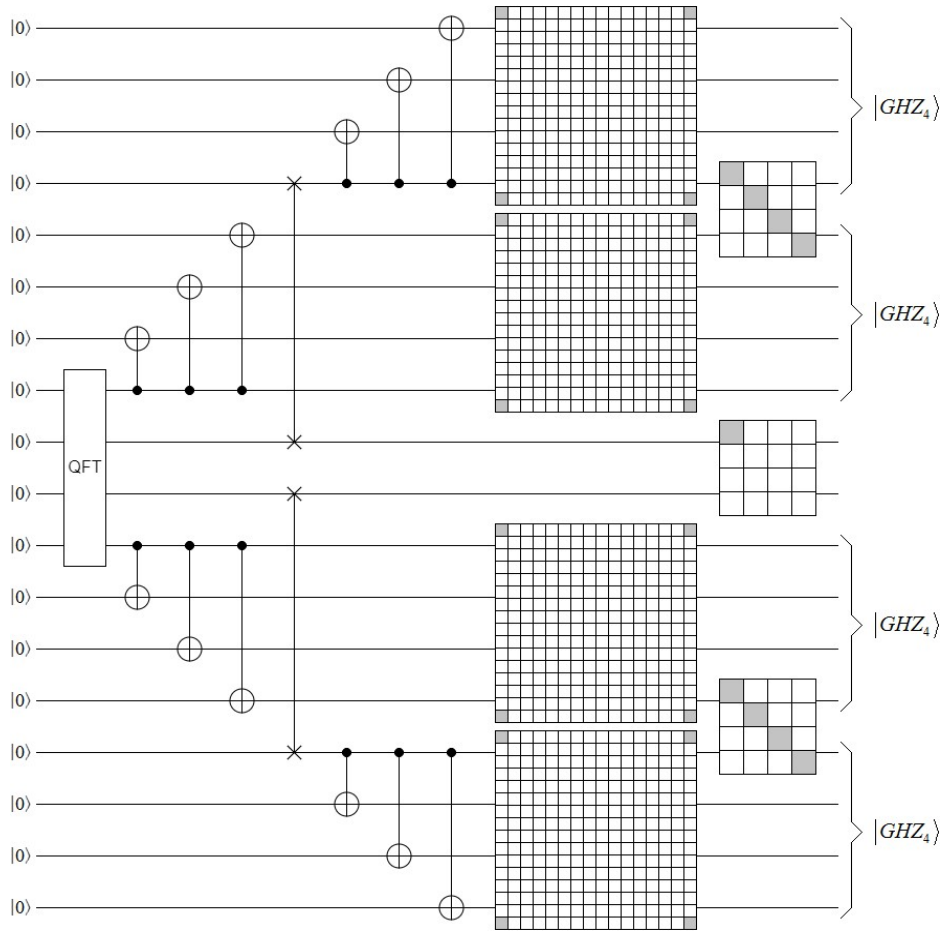

**FIGURE 4** Single source $S_4^4$ generating four independent sets of entangled states of type $|GHZ_4\rangle$. The four 8×8 central density matrices clearly show the presence of the $|GHZ_4\rangle$ states, but the two 4×4 inter-set matrices show complete independence between sets. The two SWAP gates present in the figure are used to access the two central output qubits of the 4-qubit QFT block.

in detail in Subsection 3.2.

Finally, Fig. 4 shows a single source $S_4^4$ of four sets of four entangled states of type $|GHZ_4\rangle$ each, where the four sets are completely independent of each other. In this case, a 4-qubit QFT block and two *SWAP*-type gates [5] have been used, where the *SWAP* gate for two consecutive qubits is,

$$SWAP = \begin{bmatrix} 1 & 0 & 0 & 0 \\ 0 & 0 & 1 & 0 \\ 0 & 1 & 0 & 0 \\ 0 & 0 & 0 & 1 \end{bmatrix}. \tag{15}$$

The *SWAP* gates of Eq. (15) facilitate access to the two central qubits of the QFT block. This example shows the ductility of the technique presented here, which can generate $M$ independent sets of $N$ entangled states each, opening a wide range of applications that are particularly relevant in the context of the future Quantum Internet [14-23] and which we will discuss below.

## 3 APPLICATIONS

Next, four of the main protocols involved in the arsenal of quantum communications [13] will be implemented using the entanglement parallelization technology. In particular, the selected four protocols constitute the central core of the future Quantum Internet [14-23], and in their implementation, we will use four different states, as an example without loss of generality, which respond to a strict color code in their respective histograms.

$$|\psi_1\rangle = \sqrt[4]{X}|0\rangle = \begin{bmatrix} 0.8536 + 0.3536i \\ 0.1464 - 0.3536i \end{bmatrix}, \tag{16}$$

where,

$$X = \begin{bmatrix} 0 & 1 \\ 1 & 0 \end{bmatrix}, \tag{17}$$

and,

$$\sqrt[4]{X} = \sqrt[4]{\begin{bmatrix} 0 & 1 \\ 1 & 0 \end{bmatrix}} = \begin{bmatrix} 0.8536 + 0.3536i & 0.1464 - 0.3536i \\ 0.1464 - 0.3536i & 0.8536 + 0.3536i \end{bmatrix}, \tag{18}$$

and whose density matrix is,

$$|\psi_1\rangle\langle\psi_1| = \begin{bmatrix} 0.8536 & 0.3536i \\ -0.3536i & 0.1464 \end{bmatrix}, \tag{19}$$

being the elements of its main diagonal (associated with their respective computational basis states),

$$diag_{|\psi_1\rangle} = \begin{bmatrix} 0.8536 & 0.1464 \\ |0\rangle & |1\rangle \end{bmatrix}. \tag{20}$$

Then,

$$|\psi_2\rangle = \sqrt[4]{X}|1\rangle = \begin{bmatrix} 0.1464 - 0.3536i \\ 0.8536 + 0.3536i \end{bmatrix}, \tag{21}$$

$$|\psi_2\rangle\langle\psi_2| = \begin{bmatrix} 0.1464 & -0.3536i \\ 0.3536i & 0.8536 \end{bmatrix}, \text{ and} \qquad (22)$$

$$diag_{|\psi_2\rangle} = \begin{bmatrix} 0.1464 & 0.8536 \\ |0\rangle & |1\rangle \end{bmatrix}. \qquad (23)$$

$$|\psi_3\rangle = H|0\rangle = \begin{bmatrix} 0.7071 \\ 0.7071 \end{bmatrix}, \qquad (24)$$

$$|\psi_3\rangle\langle\psi_3| = \begin{bmatrix} 0.5 & 0.5 \\ 0.5 & 0.5 \end{bmatrix}, \text{ and} \qquad (25)$$

$$diag_{|\psi_3\rangle} = \begin{bmatrix} 0.5 & 0.5 \\ |0\rangle & |1\rangle \end{bmatrix}. \qquad (26)$$

$$|\psi_4\rangle = \sqrt[4]{X}\sqrt[4]{Z}\sqrt[2]{X}|1\rangle = \begin{bmatrix} 0.8536 - 0.1464i \\ -0.3536 + 0.3536i \end{bmatrix}, \qquad (27)$$

where,

$$Z = \begin{bmatrix} 1 & 0 \\ 0 & -1 \end{bmatrix}, \qquad (28)$$

and,

$$\sqrt[4]{Z} = \sqrt[4]{\begin{bmatrix} 1 & 0 \\ 0 & -1 \end{bmatrix}} = \begin{bmatrix} 1 & 0 \\ 0 & 0.7071 + 0.7071i \end{bmatrix}, \qquad (29)$$

with,

$$\sqrt[2]{X} = \sqrt[2]{\begin{bmatrix} 0 & 1 \\ 1 & 0 \end{bmatrix}} = \begin{bmatrix} 0.5 + 0.5i & 0.5 - 0.5i \\ 0.5 - 0.5i & 0.5 + 0.5i \end{bmatrix}, \qquad (30)$$

$$|\psi_4\rangle\langle\psi_4| = \begin{bmatrix} 0.75 & -0.3536 - 0.25i \\ -0.3536 + 0.25i & 0.25 \end{bmatrix}, \text{ and} \qquad (31)$$

$$diag_{|\psi_4\rangle} = \begin{bmatrix} 0.75 & 0.25 \\ |0\rangle & |1\rangle \end{bmatrix}. \qquad (32)$$

In the four cases, i.e., Eqs. (20, 23, 26, and 32), the main diagonals are directly associated with the histograms that the IBM Q [43] platforms will deliver. This is the reason why under each element of the mentioned main diagonals of each qubit the corresponding associated CBS is in order.

Besides, the color code used in the histograms (outcomes of the IBM Q platforms) for each qubit will be: $|\psi_1\rangle$ blue, $|\psi_2\rangle$ green, $|\psi_3\rangle$ red, and $|\psi_4\rangle$ orange.

Finally, it is opportune to clarify that in all protocols, we must pay particular attention to fidelity degradation as the number of qubits involved in the quantum circuits of the different versions of the respective protocols implemented below increases, which is clearly stated in highlight in the physical platforms like the ones used here [43]. The best definition of fidelity that we can find in the literature is the following [50]: "Fidelity is the measurement of the overlap between two density matrices of theoretical and experimental quantum states obtained as output", and it can be calculated from the following formula,

$$F(\rho^T, \rho^E) = \left[ Tr\left( \sqrt{\sqrt{\rho^T} \rho^E \sqrt{\rho^T}} \right) \right]^2, \qquad (33)$$

where $F$ is the fidelity, $Tr(\bullet)$ means *the trace of a square matrix* $(\bullet)$, $\rho^T$ represents the theoretical density matrix of the qubit to be teleported (associated with the sender side), while $\rho^E$ is the experimental density matrix resulting from the platform to be used for teleportation (particularly associated with the receiver side), in this case, IBM Q [43] physical quantum processors of 5 (Belem), 7 (Oslo), and 14 (Melbourne) qubits. As we can see, it is an approach purely based on quantum tomography [5], which constitutes an excellent precedent when relying on the metric of Eq. (33). However, our experience tells us that there is a much more realistic way to calculate fidelity, which is to strictly evaluate the actual transmission process of the state in question. Specifically, the problem with this definition can be explained with a simple example.

It is important to clarify at this point that from now on, we will consider the most outstanding elements of quantum tomography of states [5], those corresponding to the main diagonal of the density matrix of the measured state, which keeps a close correspondence with the outcomes, frequencies, probabilities or components of the histogram obtained from the measurement carried out on the *z*-axis of the Bloch sphere [5] for that state on the chosen platform.

If we consider the case of the qubit $|\psi_1\rangle$, in particular, Eq. (20), and now we implement that state on any IBM Q platform [43], for example, Belem with 5 qubits, the mere fact of implementing this qubit without the intervention of any protocol for its transmission, it gives us the following values relative to its tomography:

$$tomo_{|\psi_1\rangle} = \begin{bmatrix} 0.8754 & 0.1246 \\ |0\rangle & |1\rangle \end{bmatrix}. \qquad (34)$$

The values of Eq. (34) do not coincide exactly with those theoretical values of Eq. (20). In other words, we have not transmitted anything yet, that is, we do not even use the protocol in question and there are already discrepancies due to the selected platform. Therefore, it does not make sense to calculate the fidelity between the theoretical value of the qubit to be teleported and the final reconstructed value, since there is an intermediate instance that is being completely ignored and leaves aside fundamental aspects after the evaluation of the performance of a given protocol on a given platform. For this, we must add the cumbersome fidelity calculation based on Eq. (33) due to how complicated it is to obtain the estimates of that formula [50].

Consequently, we understand that for a more realistic analysis of the transmission performance of the chosen qubit, as well as the true contribution of the platform in the degradation of that qubit, we must consider two fidelities, one associated purely to the capacity of the platform to represent a given qubit, and the other associated with the capacity of the platform to represent the chosen protocol, that is, a circuit structure more complex than that necessary to prepare a simple qubit. We will call the first one:

1. *Fidelity of representation of the state in charge of the platform* ($F_{RP}$), which is calculated on the side of Alice (sender), while the second we will call,

2. *Fidelity of the state transmission according to the platform* ($F_{TP}$), which is calculated at the end of the transmission, once we carry out the tomography of the reconstructed state on Bob's side (receiver).

If we take as an example the case of the qubit $|\psi_1\rangle$ of Eq. (16), $F_{RP}$ is calculated between the main diagonal of the theoretical value of the qubit, i.e., the elements of Eq. (20), and the tomography resulting from implementing that qubit on the chosen platform, while for the case of $F_{TP}$, this will be calculated between the tomography of the teleported qubit and that resulting from implementing that qubit on the chosen platform. This difference between the two is fundamental to assign the correct responsibilities to each part of the implementation for the reduction in the quality of the transmission.

It is important to recognize that we need to calibrate the calculation procedure for both fidelities by first creating a known state, whose theoretical elements are,

$$|\psi_A^T\rangle = \alpha_A^T|0\rangle + \beta_A^T|1\rangle = \begin{bmatrix} \alpha_A^T \\ \beta_A^T \end{bmatrix}, \tag{35}$$

where the superscript *T* means *theoretical*, the subscript *A* means *belonging to* Alice and $\alpha_A^T$ is the *projection on the spin-up* $|0\rangle$, while $\beta_A^T$ represents the *projection on the spin-down* $|1\rangle$, with $\{\alpha_A^T, \beta_A^T\} \in \mathbb{C}$, such that $|\alpha_A^T|^2 + |\beta_A^T|^2 = 1$ [5], being $|\cdot|$ the modulus of "•", and its density matrix results,

$$|\psi_A^T\rangle\langle\psi_A^T| = \begin{bmatrix} \alpha_A^T\alpha_A^{T*} & \alpha_A^T\beta_A^{T*} \\ \beta_A^T\alpha_A^{T*} & \beta_A^T\beta_A^{T*} \end{bmatrix} = \begin{bmatrix} |\alpha_A^T|^2 & \alpha_A^T\beta_A^{T*} \\ \beta_A^T\alpha_A^{T*} & |\beta_A^T|^2 \end{bmatrix}, \tag{36}$$

where (•)* means *complex conjugate of* (•). Finally, we obtain its main diagonal in the same way that we did for the cases of the four qubits to be used in all the protocols,

$$diag_{|\psi_A^T\rangle} = tomo_{|\psi_A^T\rangle} = \begin{bmatrix} |\alpha_A^T|^2 & |\beta_A^T|^2 \\ |0\rangle & |1\rangle \end{bmatrix}. \tag{37}$$

Now, we carry out the implementation of that qubit on the selected platform. This procedure is called qubit preparation [5]. Next, we carry out the tomography of the mentioned qubit, from which we recover the most outstanding elements of it, which turn out to be those of its main diagonal,

$$diag_{|\psi_A^E\rangle} = tomo_{|\psi_A^E\rangle} = \begin{bmatrix} |\alpha_A^E|^2 & |\beta_A^E|^2 \\ |0\rangle & |1\rangle \end{bmatrix}, \tag{38}$$

where the superscript *E* means *estimated*, which means that it is a physical tomography, not a theoretical one. So, the $F_{RP}$ results,

$$F_{RP} = 1 - \left\| |\alpha_A^T|^2 - |\alpha_A^E|^2 \right\| / max\left\{ |\alpha_A^T|^2, |\beta_A^T|^2 \right\} = 1 - \left\| |\beta_A^T|^2 - |\beta_A^E|^2 \right\| / max\left\{ |\alpha_A^T|^2, |\beta_A^T|^2 \right\}. \tag{39}$$

Then, we carry out the transmission of the prepared qubit, for example, employing teleportation, in such a way that on the receiver side (Bob), the result of the measurement gives us the most outstanding elements of its tomography, i.e., those of the main diagonal of the qubit density-matrix, for example,

$$diag_{|\psi_B^E\rangle} = tomo_{|\psi_B^E\rangle} = \begin{bmatrix} |\alpha_B^E|^2 & |\beta_B^E|^2 \\ |0\rangle & |1\rangle \end{bmatrix}, \tag{40}$$

where the subscript *B* means *belonging to* Bob. We then evaluate the teleportation performance from the following $F_{TP}$ equation,

$$F_{TP} = 1 - \left\| |\alpha_A^E|^2 - |\alpha_B^E|^2 \right\| / max\left\{ |\alpha_A^E|^2, |\beta_A^E|^2 \right\} = 1 - \left\| |\beta_A^E|^2 - |\beta_B^E|^2 \right\| / max\left\{ |\alpha_A^E|^2, |\beta_A^E|^2 \right\}. \tag{41}$$

In this way, we can get closer to the objective evaluation of a protocol implemented on a given platform for a qubit known to be transmitted by that protocol on that platform. In this way, we can evaluate the transmission performance of the qubit from its physical preparation in the sender (and not from theory) to its reconstruction in the receiver.

A third fidelity, which incorporates the same elements considered in the literature to find the traditional fidelity [50], is the one mentioned as *Fidelity Absolute-concerning-the-Platform* ($F_{AP}$), calculated from theory (theoretical value) to the state retrieved by Bob, i.e., between the main diagonal of the theoretical value of the qubit, and the topography of the teleported qubit,

$$F_{AP} = 1 - \left\| |\alpha_A^T|^2 - |\alpha_B^E|^2 \right\| / max\left\{ |\alpha_A^T|^2, |\beta_A^T|^2 \right\} = 1 - \left\| |\beta_A^T|^2 - |\beta_B^E|^2 \right\| / max\left\{ |\alpha_A^T|^2, |\beta_A^T|^2 \right\}. \tag{42}$$

The three fidelities ($F_{RP}$, $F_{TP}$, and $F_{AP}$) will help us draw important conclusions about the best qubit-protocol-platform combination.

### 3.1 Quantum teleportation

Teleportation [24] is the first quantum communications protocol that was created in the long list of tools belonging to the quantum communications toolbox [13]. Figure 5(a) constitutes the circuital representation of this protocol. An entangled pair $|\beta_{00}\rangle$ is created and distributed between Alice or the sender (for the top two qubits) and Bob or the receiver (exclusively associated with the lower qubit). Alice prepares the state $|\psi\rangle$ to be teleported. Both the qubit $|\psi\rangle$ to be teleported and one of the entangled particles $|\beta_{00}\rangle$ enter the Bell State Measurement (BSM) module located between the two upper qubits [5], which is made up of a *CNOT* gate, a Hadamard (H) gate, and two measurement blocks [5]. At the output of both measurement blocks, it is the classic disambiguation channel, which is represented by double lines. Two classic bits of disambiguation travel through this, which allows the selection of the gates on Bob's side (Pauli's gates: *X*-inverter, and *Z*-phase) that will lead to the reconstruction of the teleported state $|\psi\rangle$. Figures 5(b) and (c) represent the outcomes corresponding to the teleportation of the qubits $|\psi_1\rangle$ of Eq. (16) and $|\psi_3\rangle$ of Eq. (24), respectively. Although in this example we do not use a parallelization source yet, it is important for comparative purposes, since the idea is to evaluate the degradation of fidelity as the number of qubits used increases.

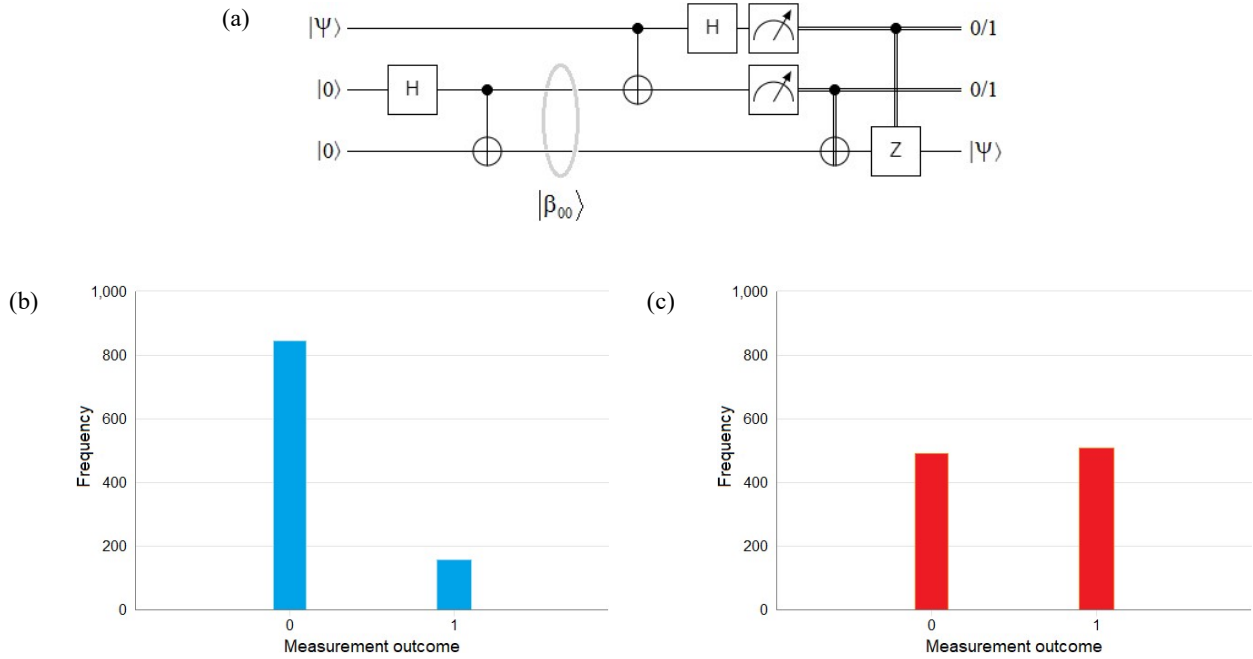

**FIGURE 5** Quantum teleportation protocol: a) quantum circuit of the protocol based on a single $|\beta_{00}\rangle$ entangled pair, b) outcomes after the teleportation of the first state type-$|\psi_1\rangle$, and then c) outcomes after the teleportation of the second state type-$|\psi_3\rangle$.

Henceforth, regardless of the number of qubits of the physical IBM Q platform used, i.e., 5 (Belem), 7 (Oslo), or 14 (Melbourne), in all cases the number of shots (attempts) will be 8192, in other words, to define each state through the respective outcomes, these will result from an average of 8192 attempts or emission of particles associated with that state.

For the teleportations of Fig. 5, we will use the IBM Q 5-qubit Belem quantum processor [43]. In the case that Alice prepares the qubit $|\psi_1\rangle$, its theoretical value arises from Eq. (2), which is,

$$frecuency_{|\psi_1^T\rangle} = \begin{bmatrix} |\alpha_1^T|^2 & |\beta_1^T|^2 \\ |0\rangle & |1\rangle \end{bmatrix} = \begin{bmatrix} 0.8536 & 0.1464 \\ |0\rangle & |1\rangle \end{bmatrix}, \tag{43}$$

while after preparation and measurement by Alice, the outcomes obtained are,

$$|\alpha_1^E|^2_A = 6141/8192 = 0.7496, \text{ and} \tag{44a}$$

$$|\beta_1^E|^2_A = 2051/8192 = 0.2504, \text{ then} \tag{44b}$$

$$outcomes_{|\psi_1^E\rangle_A} = \begin{bmatrix} |\alpha_1^E|^2_A & |\beta_1^E|^2_A \\ |0\rangle & |1\rangle \end{bmatrix} = \begin{bmatrix} 0.7496 & 0.2504 \\ |0\rangle & |1\rangle \end{bmatrix}. \tag{45}$$

As we can see, there is a discrepancy between the theoretical value of the frequency of the state $|\psi_1\rangle$ in Eq. (43) with the outcomes of the state prepared by Alice in Eq. (45), and we have not done the teleportation yet.

Once the teleportation is done, Bob gets the following outcomes,

$$|\alpha_1^E|^2_B = 5231/8192 = 0.6386, \text{ and} \tag{46a}$$

$$|\beta_1^E|^2_B = 2961/8192 = 0.3614, \text{ then} \tag{46b}$$

$$outcomes_{|\psi_1^E\rangle_B} = \begin{bmatrix} |\alpha_1^E|^2_B & |\beta_1^E|^2_B \\ |0\rangle & |1\rangle \end{bmatrix} = \begin{bmatrix} 0.6386 & 0.3614 \\ |0\rangle & |1\rangle \end{bmatrix}. \tag{47}$$

It is important to clarify at this point that Alice prepares the state $|\psi_1\rangle$ and measures it without teleporting it, to assess the degree of degradation due to the decoherence of the processor used, regardless of the teleportation. In a second and independent process, Alice prepares the same state again and teleports it away. In the second case, the only measurement carried out by Alice corresponds to the BSM necessary as an essential part of any teleportation. In summary, this two-step procedure gives us an idea of the error propagation that is coming due to the processor used.

The fidelities associated with the outcomes of Fig. 5(b) for the case of the qubit $|\psi_1\rangle$ are:

$$F_{RP,|\psi_1\rangle} = 1 - \left||\alpha_1^T|^2_A - |\alpha_1^E|^2_A\right|/max\left\{|\alpha_1^T|^2_A, |\beta_1^T|^2_A\right\} = 1 - \left||\beta_1^T|^2_A - |\beta_1^E|^2_A\right|/max\left\{|\alpha_1^T|^2_A, |\beta_1^T|^2_A\right\}$$
$$= 1 - |0.8536 - 0.7496|/0.8536 = 1 - |0.1464 - 0.2504|/0.8536 = 0.8782, \tag{48}$$

$$F_{TP,|\psi_1\rangle} = 1 - \left||\alpha_1^E|^2_A - |\alpha_1^E|^2_B\right|/max\left\{|\alpha_1^E|^2_A, |\beta_1^E|^2_A\right\} = 1 - \left||\beta_1^E|^2_A - |\beta_1^E|^2_B\right|/max\left\{|\alpha_1^E|^2_A, |\beta_1^E|^2_A\right\}$$
$$= 1 - |0.7496 - 0.6386|/0.7496 = 1 - |0.2504 - 0.3614|/0.7496 = 0.8519, \tag{49}$$

while the absolute fidelity on the platform is:

$$F_{AP,|\psi_1\rangle} = 1 - \left\| |\alpha_1^T|^2 \Big|_A - |\alpha_1^E|^2 \Big|_B \right\| / max\left\{ |\alpha_1^T|^2 \Big|_A, |\beta_1^T|^2 \Big|_A \right\} = 1 - \left\| |\beta_1^T|^2 \Big|_A - |\beta_1^E|^2 \Big|_B \right\| / max\left\{ |\alpha_1^T|^2 \Big|_A, |\beta_1^T|^2 \Big|_A \right\}$$
$$= 1 - |0.8536 - 0.7458|/0.8536 = 1 - |0.1464 - 0.2542|/0.8536 = 0.8737. \tag{50}$$

The absolute fidelity on the platform, which incorporates the same elements considered in the literature to find the traditional fidelity [50], gives us values similar to the $F_{RP}$ and not to the $F_{TP}$, which tells us of a transmission performance superior to the real one, closer to the implementation of the qubit on the platform than to its transmission, which leaves much to be desired since the protocol circuit of Fig. 5 is much more complex than that necessary to prepare the qubit $|\psi_1\rangle$. This tells us that the fidelity commonly used in the literature [50] gives us better performance than the real one.

For the case of state $|\psi_3\rangle$ of Eq. (24), its theoretical value is,

$$frecuency_{|\psi_3^T\rangle} = \begin{bmatrix} |\alpha_3^T|^2 & |\beta_3^T|^2 \\ |0\rangle & |1\rangle \end{bmatrix} = \begin{bmatrix} 0.5 & 0.5 \\ |0\rangle & |1\rangle \end{bmatrix}, \tag{51}$$

while after preparation and measurement by Alice, the outcomes obtained are,

$$|\alpha_3^E|^2 \Big|_A = 4040/8192 = 0.4932, \text{ and} \tag{52a}$$

$$|\beta_3^E|^2 \Big|_A = 4152/8192 = 0.5068, \text{ then} \tag{52b}$$

$$outcomes_{|\psi_3^E\rangle_A} = \begin{bmatrix} |\alpha_3^E|^2 \Big|_A & |\beta_3^E|^2 \Big|_A \\ |0\rangle & |1\rangle \end{bmatrix} = \begin{bmatrix} 0.4932 & 0.5068 \\ |0\rangle & |1\rangle \end{bmatrix}. \tag{53}$$

Comparing Eqs. (51), and (53), we can observe a degradation similar to the case seen between Eqs. (43), and (45), for the case of the state $|\psi_1\rangle$.

Once the teleportation is done, Bob gets the following outcomes,

$$|\alpha_3^E|^2 \Big|_B = 4005/8192 = 0.4889, \text{ and} \tag{54a}$$

$$|\beta_3^E|^2 \Big|_B = 4187/8192 = 0.5111, \text{ then} \tag{54b}$$

$$outcomes_{|\psi_3^E\rangle_B} = \begin{bmatrix} |\alpha_3^E|^2 \Big|_B & |\beta_3^E|^2 \Big|_B \\ |0\rangle & |1\rangle \end{bmatrix} = \begin{bmatrix} 0.4889 & 0.5111 \\ |0\rangle & |1\rangle \end{bmatrix}. \tag{55}$$

The fidelities associated with the outcomes of Fig. 5(c) for the case of the qubit $|\psi_3\rangle$ are:

$$F_{RP,|\psi_3\rangle} = 1 - \left\| |\alpha_3^T|^2 \Big|_A - |\alpha_3^E|^2 \Big|_A \right\| / max\left\{ |\alpha_3^T|^2 \Big|_A, |\beta_3^T|^2 \Big|_A \right\} = 1 - \left\| |\beta_3^T|^2 \Big|_A - |\beta_3^E|^2 \Big|_A \right\| / max\left\{ |\alpha_3^T|^2 \Big|_A, |\beta_3^T|^2 \Big|_A \right\}$$
$$= 1 - |0.5 - 0.4932|/0.5 = 1 - |0.5 - 0.5068|/0.5 = 0.9865, \tag{56}$$

$$F_{TP,|\psi_3\rangle} = 1 - \left\| |\alpha_3^E|^2 \Big|_A - |\alpha_3^E|^2 \Big|_B \right\| / max\left\{ |\alpha_3^E|^2 \Big|_A, |\beta_3^E|^2 \Big|_A \right\} = 1 - \left\| |\beta_3^E|^2 \Big|_A - |\beta_3^E|^2 \Big|_B \right\| / max\left\{ |\alpha_3^E|^2 \Big|_A, |\beta_3^E|^2 \Big|_A \right\}$$
$$= 1 - |0.4932 - 0.4889|/0.5068 = 1 - |0.5068 - 0.5111|/0.5068 = 0.9913, \tag{57}$$

while the absolute fidelity on the platform is:

$$F_{AP,|\psi_3\rangle} = 1 - \left\| |\alpha_3^T|^2_A - |\alpha_3^E|^2_B \right\| / max\left\{ |\alpha_3^T|^2_A, |\beta_3^T|^2_A \right\} = 1 - \left\| |\beta_3^T|^2_A - |\beta_3^E|^2_B \right\| / max\left\{ |\alpha_3^T|^2_A, |\beta_3^T|^2_A \right\} \qquad (58)$$
$$= 1 - |0.5 - 0.4889|/0.5 = 1 - |0.5 - 0.5111|/0.5 = 0.9778.$$

The diagonal state $|\psi_3\rangle = |+\rangle$ of Eq. (24) usually gives better results on all kinds of platforms than a qubit of the $|\psi_1\rangle$-type, i.e., better than a more generic qubit like $|\psi_1\rangle$, even in the case of an optical table [25-27].

Figure 6(a) represents a simultaneous teleportation of qubits $|\psi_1\rangle$ and $|\psi_3\rangle$ thanks to a single source of two sets of entangled particles of $|\beta_{00}\rangle$-type. The two SWAP gates present in the figure are used to access the two respective output qubits of the 2-qubit QFT block (i.e., a $2^2 \times 2^2$ matrix). For these experiments, we use the 7-qubit Oslo IBM Q processor [43]. The theoretical frequencies of both qubits remain those of Eqs. (43), and (51). Therefore, when sender #1 prepares and measures its qubit $|\psi_1\rangle$, the outcomes obtained are:

$$|\alpha_1^E|^2_A = 6058/8192 = 0.7395 \text{, and} \qquad (59a)$$

$$|\beta_1^E|^2_A = 2134/8192 = 0.2605 \text{, then} \qquad (59b)$$

$$outcomes_{|\psi_1^E\rangle_A} = \begin{bmatrix} |\alpha_1^E|^2_A & |\beta_1^E|^2_A \\ |0\rangle & |1\rangle \end{bmatrix} = \begin{bmatrix} 0.7395 & 0.2605 \\ |0\rangle & |1\rangle \end{bmatrix}. \qquad (60)$$

The discrepancy between the theoretical value of the frequency of the state in Eq. (43) with the outcomes of the state prepared by sender #1 in Eq. (60) is very similar to that found for the same case in Fig. 5(b) between Eqs. (43) and (45) working with the 5-qubit Belem IBM Q processor [43].

Once the teleportation is done, receiver #1 gets the following outcomes,

$$|\alpha_1^E|^2_B = 5768/8192 = 0.7041 \text{, and} \qquad (61a)$$

$$|\beta_1^E|^2_B = 2424/8192 = 0.2959 \text{, then} \qquad (61b)$$

$$outcomes_{|\psi_1^E\rangle_B} = \begin{bmatrix} |\alpha_1^E|^2_B & |\beta_1^E|^2_B \\ |0\rangle & |1\rangle \end{bmatrix} = \begin{bmatrix} 0.7041 & 0.2959 \\ |0\rangle & |1\rangle \end{bmatrix}. \qquad (62)$$

The fidelities associated with the outcomes of Fig. 6(b) for the case of the qubit $|\psi_1\rangle$ are:

$$F_{RP,|\psi_1\rangle} = 1 - \left\| |\alpha_1^T|^2_A - |\alpha_1^E|^2_A \right\| / max\left\{ |\alpha_1^T|^2_A, |\beta_1^T|^2_A \right\} = 1 - \left\| |\beta_1^T|^2_A - |\beta_1^E|^2_A \right\| / max\left\{ |\alpha_1^T|^2_A, |\beta_1^T|^2_A \right\} \qquad (63)$$
$$= 1 - |0.8536 - 0.7395|/0.8536 = 1 - |0.1464 - 0.2605|/0.8536 = 0.8664,$$

$$F_{TP,|\psi_1\rangle} = 1 - \left\| |\alpha_1^E|^2_A - |\alpha_1^E|^2_B \right\| / max\left\{ |\alpha_1^E|^2_A, |\beta_1^E|^2_A \right\} = 1 - \left\| |\beta_1^E|^2_A - |\beta_1^E|^2_B \right\| / max\left\{ |\alpha_1^E|^2_A, |\beta_1^E|^2_A \right\} \qquad (64)$$
$$= 1 - |0.7395 - 0.7041|/0.7395 = 1 - |0.2605 - 0.2959|/0.7395 = 0.9522,$$

while the absolute fidelity on the platform is:

$$F_{AP,|\psi_1\rangle} = 1 - \left| |\alpha_1^T|^2_A - |\alpha_1^E|^2_B \right| / max\left\{ |\alpha_1^T|^2_A, |\beta_1^T|^2_A \right\} = 1 - \left| |\beta_1^T|^2_A - |\beta_1^E|^2_B \right| / max\left\{ |\alpha_1^T|^2_A, |\beta_1^T|^2_A \right\}$$
$$= 1 - |0.8536 - 0.7041|/0.8536 = 1 - |0.1464 - 0.2959|/0.8536 = 0.9142.$$
(65)

As in the case of the configuration in Fig. 5, where the fidelities of Eqs. (48), (49), and (50) indicate that the traditional fidelity used in the literature [50] gives superior performance to the real one, the fidelities obtained here through Eqs. (63), (64), and (65), testify similar behavior, even with a different quantum processor.

Now, when sender #3 prepares and measures its qubit $|\psi_3\rangle$, the outcomes obtained are:

$$\left|\alpha_3^E\right|^2_A = 3917/8192 = 0.4781, \text{ and} \tag{66a}$$

$$\left|\beta_3^E\right|^2_A = 4275/8192 = 0.5219, \text{ then} \tag{66b}$$

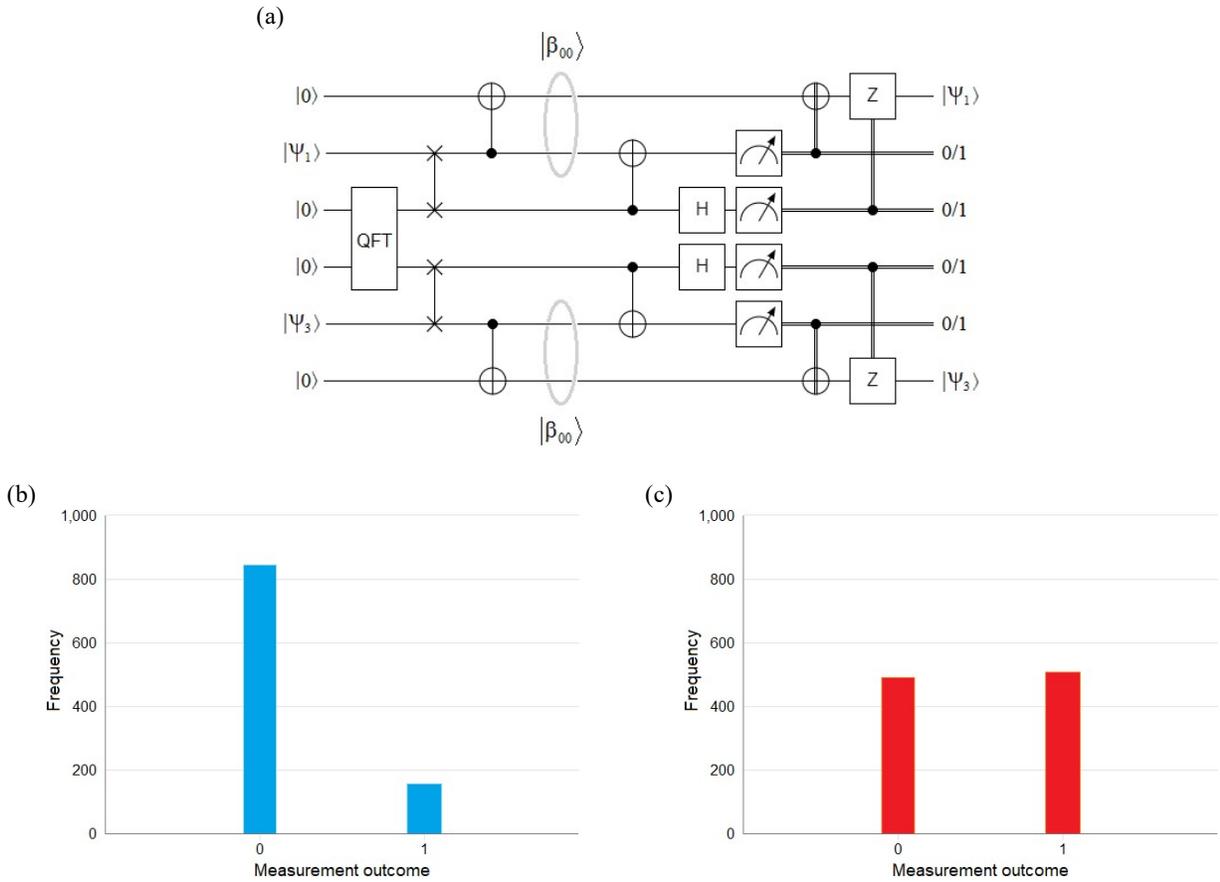

**FIGURE 6** Two simultaneous teleportations thanks to the parallel entanglement technology: a) quantum circuit of the protocol with a 2-qubit QFT block (which generates two independent pairs of entangled particles $|\beta_{00}\rangle$), and two SWAP gates, which facilitate the wiring between the two states to be teleported $\{|\psi_1\rangle, |\psi_3\rangle\}$ and the respective outputs of the QFT block, b) outcomes obtained after the teleportation of the $|\psi_1\rangle$ state, and c) outcomes obtained after the teleportation of the $|\psi_3\rangle$ state.

$$outcomes_{|\psi_3^E\rangle_A} = \begin{bmatrix} |\alpha_3^E|^2_A & |\beta_3^E|^2_A \\ |0\rangle & |1\rangle \end{bmatrix} = \begin{bmatrix} 0.4781 & 0.5219 \\ |0\rangle & |1\rangle \end{bmatrix}. \tag{67}$$

Comparing Eqs. (51), and (57), we can observe a similar degradation to the case seen between Eqs. (51), and (53), for the case of the same state $|\psi_1\rangle$ of Fig. 5(c).

Once the teleportation is done, receiver #3 gets the following outcomes,

$$|\alpha_3^E|^2_B = 4026/8192 = 0.4914, \text{ and} \tag{68a}$$

$$|\beta_3^E|^2_B = 4166/8192 = 0.5086, \text{ then} \tag{68b}$$

$$outcomes_{|\psi_3^E\rangle_B} = \begin{bmatrix} |\alpha_3^E|^2_B & |\beta_3^E|^2_B \\ |0\rangle & |1\rangle \end{bmatrix} = \begin{bmatrix} 0.4914 & 0.5086 \\ |0\rangle & |1\rangle \end{bmatrix}. \tag{69}$$

In Eq. (69), we can observe a greater degradation of the outcomes than in the case of Eq. (55) corresponding to Fig. 5(c). The fidelities associated with the outcomes of Fig. 6(c) for the case of the qubit $|\psi_3\rangle$ are:

$$F_{RP,|\psi_3\rangle} = 1 - \left||\alpha_3^T|^2_A - |\alpha_3^E|^2_A\right|/max\{|\alpha_3^T|^2_A, |\beta_3^T|^2_A\} = 1 - \left||\beta_3^T|^2_A - |\beta_3^E|^2_A\right|/max\{|\alpha_3^T|^2_A, |\beta_3^T|^2_A\}$$
$$= 1 - |0.5 - 0.4781|/0.5 = 1 - |0.5 - 0.5219|/0.5 = 0.9562, \tag{70}$$

$$F_{TP,|\psi_3\rangle} = 1 - \left||\alpha_3^E|^2_A - |\alpha_3^E|^2_B\right|/max\{|\alpha_3^E|^2_A, |\beta_3^E|^2_A\} = 1 - \left||\beta_3^E|^2_A - |\beta_3^E|^2_B\right|/max\{|\alpha_3^E|^2_A, |\beta_3^E|^2_A\}$$
$$= 1 - |0.4781 - 0.4914|/0.5219 = 1 - |0.5219 - 0.5086|/0.5219 = 0.9416, \tag{71}$$

while the absolute fidelity on the platform is:

$$F_{AP,|\psi_3\rangle} = 1 - \left||\alpha_3^T|^2_A - |\alpha_3^E|^2_B\right|/max\{|\alpha_3^T|^2_A, |\beta_3^T|^2_A\} = 1 - \left||\beta_3^T|^2_A - |\beta_3^E|^2_B\right|/max\{|\alpha_3^T|^2_A, |\beta_3^T|^2_A\}$$
$$= 1 - |0.5 - 0.4914|/0.5 = 1 - |0.5 - 0.5086|/0.5 = 0.9646. \tag{72}$$

Comparing the fidelities of Eqs. (63), (64), and (65), for the case of Fig. 6(b), with those of Eqs. (70), (71), and (72), for the case of Fig. 6(c), it is evident the degradation of the first one, i.e., the teleportation of the qubit $|\psi_1\rangle$ has worse results than those obtained for the case of the qubit $|\psi_3\rangle$, which represents a similar situation to that obtained for the sequential teleportations of Fig. 5(b) and (c). Regarding the results in general, although the 7-qubit IBM Q Oslo processor is a relatively new unit, its intensive and uninterrupted use gives rise to noticeable decoherence, which is reflected in fidelities of less than one.

Finally, Fig. 7(a) represents four simultaneous teleportations thanks to a 4-qubit QFT block. Figures 7(b, c, d, and e) represent the outcomes resulting from the teleportations of qubits $|\psi_1\rangle$, $|\psi_2\rangle$, $|\psi_3\rangle$, and $|\psi_4\rangle$, respectively. In this experiment, a 14-qubit quantum processor from the IBM Q family called Melbourne [43] is used. The fidelities involved in the four teleportations can be seen in Table 1. As far as qubits $|\psi_1\rangle$ and $|\psi_3\rangle$ are concerned, there is a notable degradation of fidelities concerning the experiments in Figs. 5 and 6, which is due to three fundamental factors:

1) the increase in the number of qubits and gates of the circuit of Fig. 7(a) respect to those of Figs. 5(a) and 6(a),
2) the Melbourne processor is older than Belem and Oslo, and therefore has more decoherence [5, 43], and
3) the complexity of the 4-qubit QFT block (i.e., a $2^4 \times 2^4$ matrix) is higher than its 2-qubit counterpart.

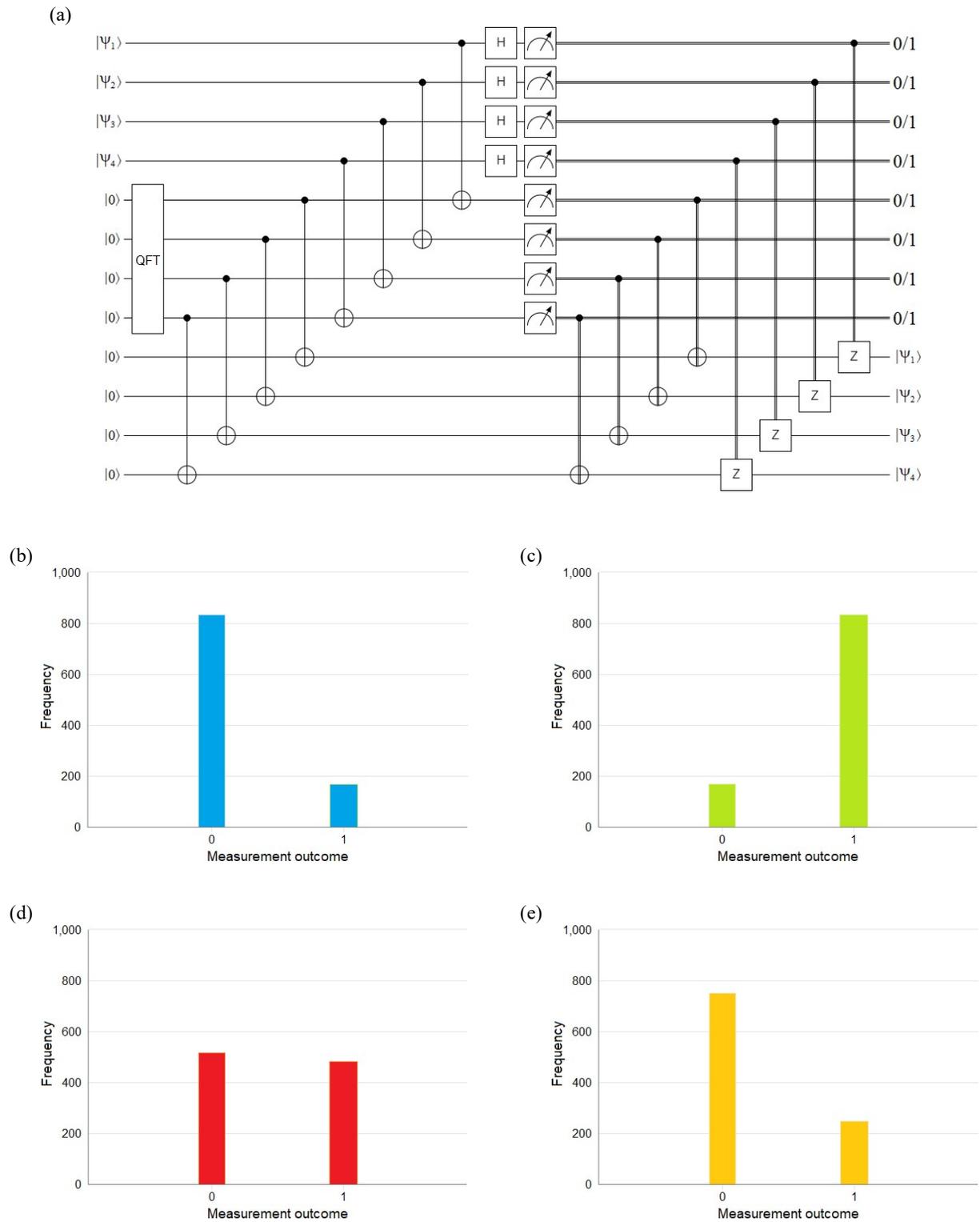

**FIGURE 7** Four simultaneous teleportations thanks to a 4-qubit QFT block: a) quantum circuit, b) outcomes of $|\psi_1\rangle$, c) outcomes of $|\psi_2\rangle$, d) outcomes of $|\psi_3\rangle$, and e) outcomes of $|\psi_4\rangle$.

**TABLE 1** Fidelities associated with the simultaneous teleportation of the four states.

| Qubits | $F_{RP}$ | $F_{TP}$ | $F_{AP}$ |
|---|---|---|---|
| $|\psi_1\rangle$ | 0.8667 | 0.9264 | 0.8903 |
| $|\psi_2\rangle$ | 0.8612 | 0.9298 | 0.9114 |
| $|\psi_3\rangle$ | 0.9587 | 0.9415 | 0.9228 |
| $|\psi_4\rangle$ | 0.9018 | 0.9125 | 0.9043 |

## 3.2 Quantum secret sharing (QSS)

Among the typical applications of states of the $|GHZ_N\rangle$-type, we can mention their participation in experiments related to Bell's theorem [53], and in the QSS protocol [31]. Figure 8(a) represents this protocol, which is a core member of the quantum cryptography family of protocols [30]. As can be seen, it is a configuration quite similar to that of the quantum teleportation protocol [24], but with more gates than this one. For this experiment, we sequentially performed the teleportation of states $|\psi_1\rangle$ and $|\psi_3\rangle$ in the 5-qubit Belem processor of IBM Q [43]. Figures 8(b), and (c) represent, respectively, the final outcomes obtained in each case.

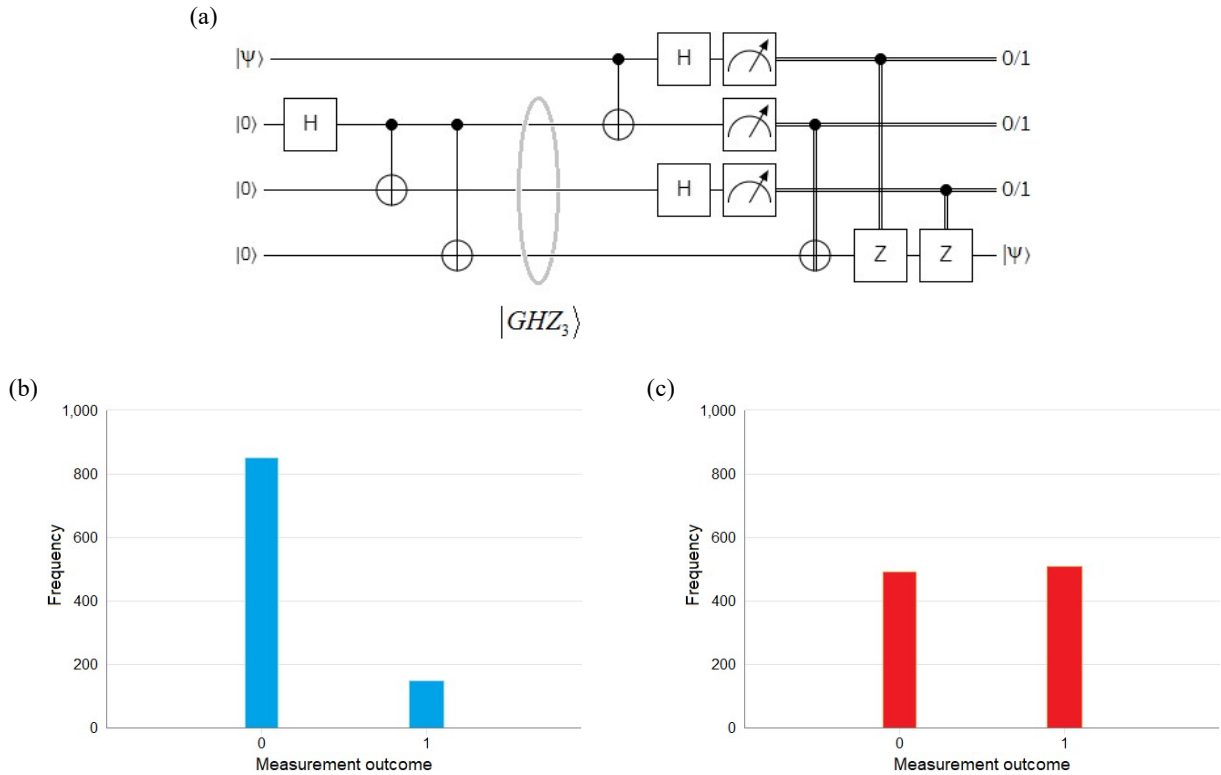

**FIGURE 8** Quantum secret sharing (QSS) protocol: a) quantum circuit of the protocol based on a single $|GHZ_3\rangle$ entangled state, b) outcomes after the quantum secret sharing of the first state type $|\psi_1\rangle$, and then c) outcomes after the quantum secret sharing of the second state type $|\psi_3\rangle$.

**TABLE 2** Fidelities associated with the sequential transmission of states $|\psi_1\rangle$ and $|\psi_3\rangle$.

| Qubits | $F_{RP}$ | $F_{TP}$ | $F_{AP}$ |
|---|---|---|---|
| $|\psi_1\rangle$ | 0.8212 | 0.9280 | 0.8932 |
| $|\psi_3\rangle$ | 0.9712 | 0.9401 | 0.9689 |

Table 2 shows a marked degradation between both sets of fidelities corresponding to this experiment, i.e., for states $|\psi_1\rangle$ and $|\psi_3\rangle$. Moreover, the second set of fidelities is very similar to its counterpart obtained in the single teleportation of Fig. 5(c) for the case of the qubit $|\psi_3\rangle$, however, the first set of fidelities is very different from its counterpart obtained in the single teleportation of Fig. 5(b) for the case of the qubit $|\psi_1\rangle$.

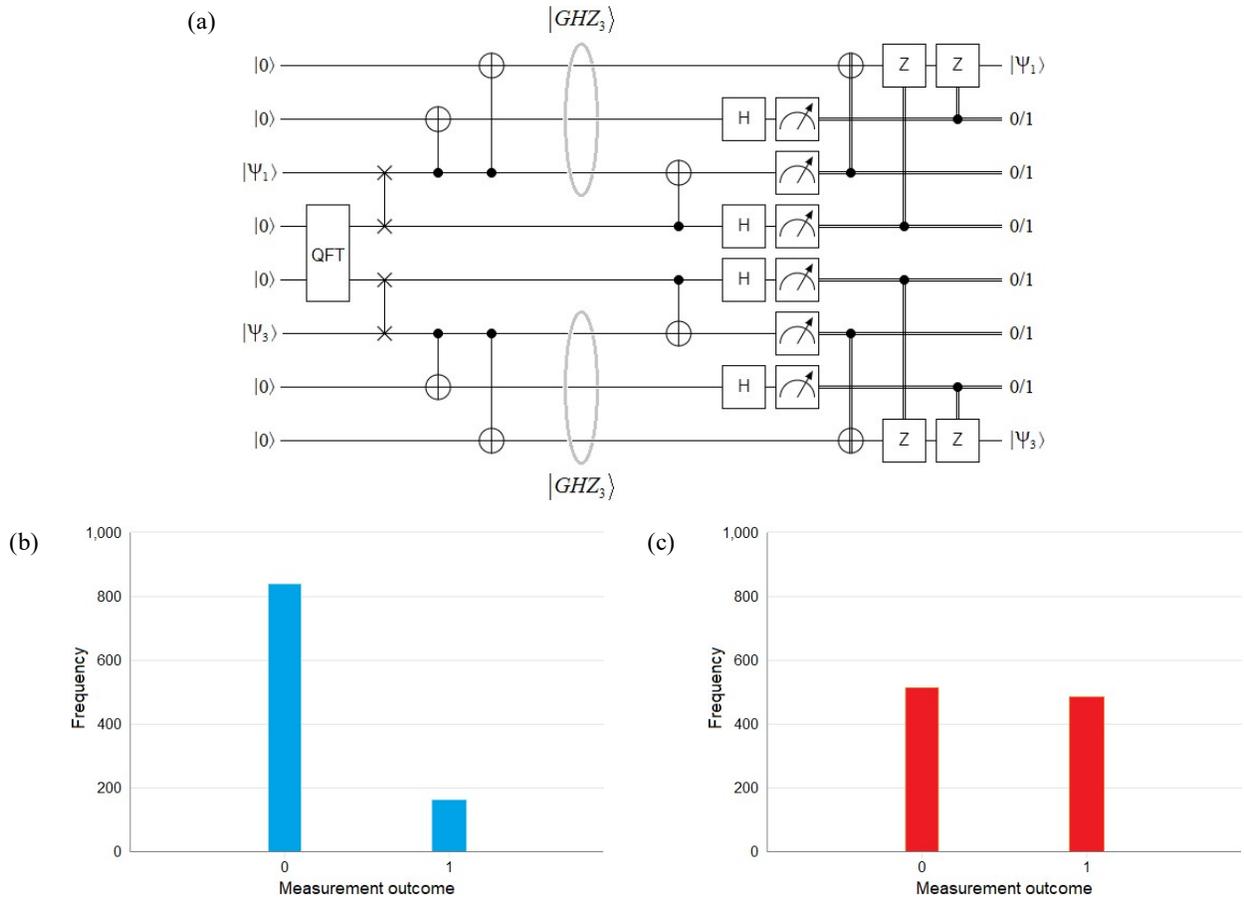

**FIGURE 9** Two simultaneous quantum secret sharing thanks to the parallel entanglement technology: a) quantum circuit of the protocol with a 2-qubit QFT block (which generates two independent triads of entangled particles $|GHZ_3\rangle$), and two SWAP gates, which facilitate the wiring between the two states to be quantum-secret-shared $\{|\psi_1\rangle,|\psi_3\rangle\}$ and the respective outputs of the QFT block, b) outcomes obtained after the quantum secret sharing of the $|\psi_1\rangle$ state, and c) outcomes obtained after the quantum secret sharing of the $|\psi_3\rangle$ state.

Figure 9(a) represents the simultaneous version of the configuration of Fig. 8(a) for the same qubits to be quantum-secret-shared, i.e., $|\psi_1\rangle$ and $|\psi_3\rangle$, thanks to a single source of two sets of $|GHZ_3\rangle$ entangled particles based on a 2-qubit QFT block. In this experiment, the 14-qubit IBM Q Melbourne processor is used. Figures 9(b) and (c) show the outcomes obtained when $|\psi_1\rangle$ and $|\psi_3\rangle$ are quantum-secret-shared, respectively. Table 3 represents the fidelities associated with this experiment. The results give a worse performance than in the case of the experiment in Fig. 8.

**TABLE 3** Fidelities associated with the simultaneous teleportation of states $|\psi_1\rangle$ and $|\psi_3\rangle$.

| Qubits | $F_{RP}$ | $F_{TP}$ | $F_{AP}$ |
|---|---|---|---|
| $|\psi_1\rangle$ | 0.8209 | 0.9239 | 0.8914 |
| $|\psi_3\rangle$ | 0.9641 | 0.9273 | 0.9632 |

Through all the experiments carried out up to now, it is extremely evident that the $F_{AP}$ fidelity is always degraded for the case of the transmission of the first qubit, which, by experiments carried out in an optical table, we know that the outcomes should be worse, that is, our results coincide with those of literature, which tells us that a state $|\psi_3\rangle = |+\rangle$ should be transmitted as better fidelity than a state $|\psi_1\rangle$, which is more generic. On the other hand, when the number of gates is increased from the configuration of Fig. 8(a) to that of Fig. 9(a), the fidelity results should be worse given the involvement of a series of errors associated with the increase of gates such as bit-flip, phase-flip, and bit-phase-flip [5]. Therefore, the results are in the right direction. Furthermore, Melbourne is older than Oslo, and therefore its decoherence is greater. We must also take into account the intensity of continuous use of a higher qubit processor, which is more required by users.

### 3.3 Bidirectional teleportation

Figure 10(a) shows a version of the protocol for the bidirectional teleportation of two qubits at the same time [54] based on a QFT block. In Fig. 10(a), three sectors can be observed: a) Alice's sector (pink), where she is the sender and the receiver at the same time, b) Bob's sector (blue), where he is the sender and the receiver at the same time too, and c) Charlie's sector (green), which generates and distributes two disjoint pairs of entangled particles of the $|\beta_{00}\rangle$-type, thanks to a 2-qubit QFT block, which crosses its outputs to give rise to the bi-directionality of the protocol. Figures 10(b), and (c) represent the outcomes for the bidirectional and simultaneous teleportation of the $|\psi_1\rangle$, and $|\psi_3\rangle$ qubits. The quantum platform employed in this experiment is Oslo from IBM Q [43], with 7 qubits.

**TABLE 4** Fidelities associated with the bidirectional and simultaneous teleportation of states $|\psi_1\rangle$ and $|\psi_3\rangle$.

| Qubits | $F_{RP}$ | $F_{TP}$ | $F_{AP}$ |
|---|---|---|---|
| $|\psi_1\rangle$ | 0.8643 | 0.9469 | 0.9174 |
| $|\psi_3\rangle$ | 0.9443 | 0.9465 | 0.9678 |

Table 4 shows the fidelities associated with this experiment for both qubits. The results obtained in Table 4 are very similar to those obtained for the experiment of Fig. 6, which was carried out with the same number of qubits, the same number of gates, and the same processor. Then, the expected results have been obtained. As in the previous cases, we continue to observe a total coincidence in the results concerning those of literature [25-29], since the $|\psi_3\rangle$ qubit has a better teleportation performance than $|\psi_1\rangle$.

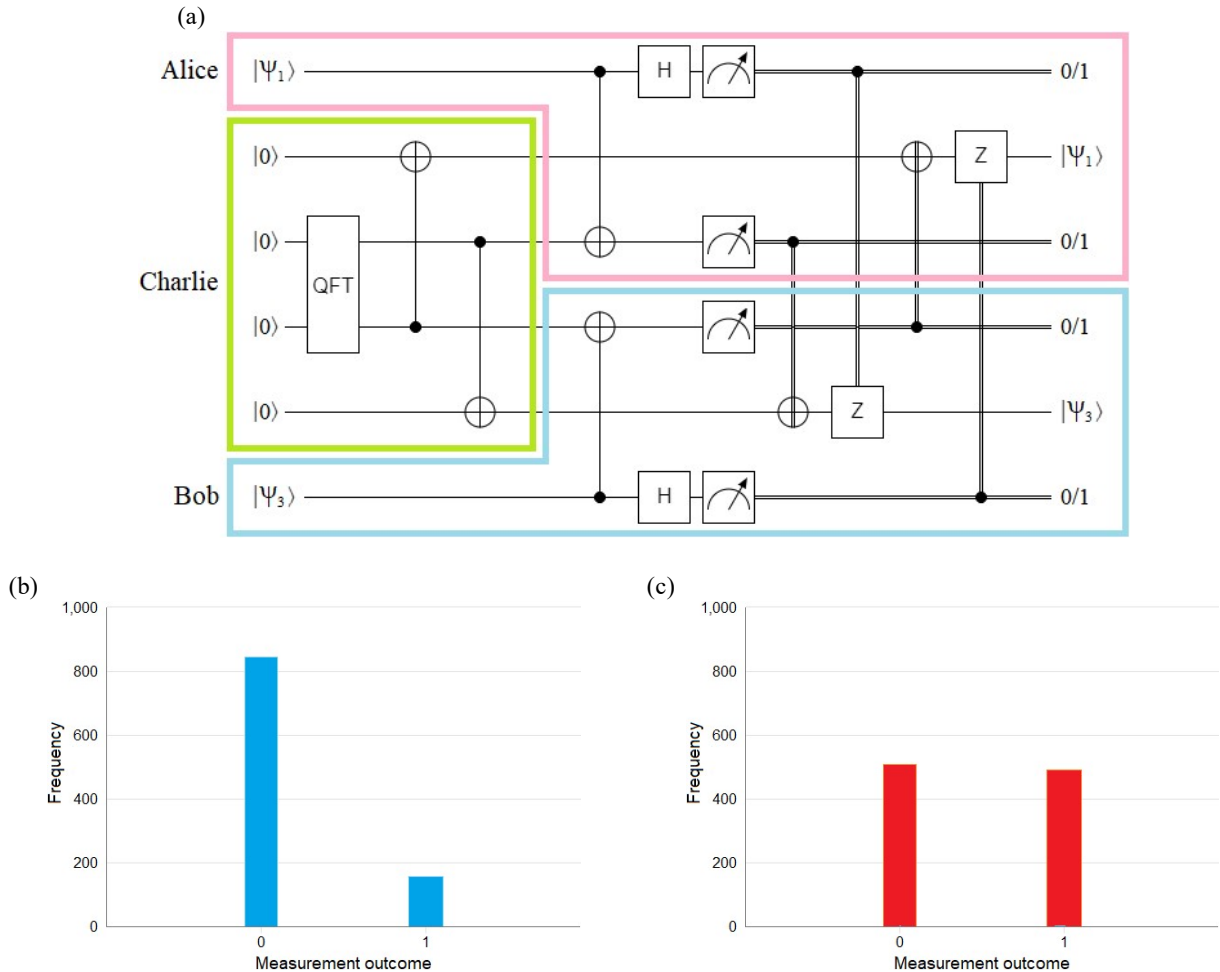

**FIGURE 10** Bidirectional teleportation protocol based on entanglement parallelization: a) quantum circuit, where Charlie (green) generates and distributes two independent pairs of entangled particles of type $|\beta_{00}\rangle$ thanks to a 2-qubit QFT block, Alice prepares the state $|\psi_1\rangle$ to be teleported to Bob, and Bob prepares the state $|\psi_3\rangle$ to be teleported to Alice, b) outcomes of the $|\psi_1\rangle$ state reconstructed on Bob side and c) outcomes of the $|\psi_3\rangle$ state reconstructed on Alice side.

### 3.4 Quantum repeaters

In all the protocols seen so far, as well as in those that are studied below, we always have two types of channels: the quantum, and the classical. The first one results from the distribution of entangled particles, which currently consist almost exclusively of photons (and therefore this channel turns out to be optical), while the second one is a digital type signal, which can travel electromagnetically or optically, although in the latter case it does not do so using entangled photons. The quantum channel can then be established over fiber optic lines or the vacuum of space, in the latter case through the intervention of satellites. The transmission of entangled photons through fiber optic lines is associated with some technical difficulties:

a) the best available fibers have a loss by absorption of 0.15 dB/km at the optimal wavelength [55],
b) light in fiber optics travels at 2/3 of the speed of light in the vacuum [55], and
c) there are spurious diffractions in the optical fiber, whereby part of the transmission energy of the entangled photons is lost [55]. Therefore, quantum repeaters [56-59] must be used every 50 km.

On the other hand, a free space link is particularly affected by the interruption of the line of sight between an optical ground station and a satellite or between satellites, where in both cases the line of sight can be interrupted by the interposition of the Earth. In both cases, the problem is solved by using satellite relays [55, 60-62]. The current trend consists of the intensive use of satellite relays over terrestrial fiber optic lines [55]. However, we present below two techniques for the construction of quantum repeaters, one oriented to terrestrial fiber optic lines and known as *entanglement swapping* [63-69], and the other particularly useful for the case of satellite relays and which we will call here *forward cascade* [70].

*3.4.1 Entanglement swapping*
The quantum repeaters used in fiber optic lines to extend the reach of the quantum channel in QKD protocols based on entangled pairs [37] are implemented by a technique known as entanglement swapping [39, 40, 63-69], which requires intensive use of quantum memories [70]. These work as temporary buffers generating delays necessary to synchronize actions between the transmitter (Alice) and the receiver (Bob).

Figure 11(a) shows a typical implementation of the entanglement swapping protocol [39, 40, 63-69], while Fig. 11(b) represents a modification based on the entanglement parallelization, in which two independent sources for the generation and distribution of entangled pairs of the $|\beta_{00}\rangle$- type is replaced by a single source generating two independent sets of entangled pairs of the same type, albeit from a 2-qubit QFT block.

In both versions, we will carry out the implementations in the 5-qubit Belem IBM Q processor [43], being $|\psi_1\rangle$ the only qubit to be transmitted in both cases for comparison purposes. Figure 11(c) represents the outcomes for the version of Fig. 11(a), while Fig. 11(d) does the same, but for the version of Fig. 11(b). Table 5 shows that the performance of the version in Fig. 11(b) is slightly superior to its counterpart in the original version in Fig. 11(a). Besides, the three fidelities give worse values to the simple teleportation of Fig. 5(a), but better to the double teleportation of Fig. 6(a), and we associate these results with the number of gates involved.

**TABLE 5** Fidelities associated with the entanglement swapping protocols when they transmit the state $|\psi_1\rangle$.

| Version | $F_{RP}$ | $F_{TP}$ | $F_{AP}$ |
|---|---|---|---|
| Figs. 11(a), and (c) | 0.8535 | 0.8478 | 0.8413 |
| Figs. 11(b), and (d) | 0.8562 | 0.8474 | 0.8536 |

*3.4.2 Forward cascade*
Although modern classical telecommunication relies on optical fibers, the direct transmission of photons through fibers is not practical for quantum communication over global distances, because losses are too high [60]. As we have mentioned before, the best available fibers have a loss of 0.15 dB/km at the optimal wavelength [55]. This means, for example, that the time to distribute one entangled photon pair over 2000 km with a 1 GHz source exceeds the age of the universe [60]. For this reason, the current trend in quantum repeaters for the Quantum Internet is mainly focused on satellite links [55, 60-62].

Figures 12-I present four experiments involving quantum repeaters under the forward cascade technique [55]. In the four mentioned experiments, we will transmit the qubit $|\psi_1\rangle$, and in all cases, we will use the 7-qubit Oslo IBM Q processor for comparison purposes. The idea is to carry out a comparison of different sources generating entangled particles from the point of view of their performance, which is used in quantum repeaters that employ the forward cascade technique.

We start with Fig. 12-I (a), which represents a teleportation (upper left side of the figure) that extends its transmission range thanks to the use of another teleportation that acts as a quantum repeater (highlighted in yellow) working in forward cascade mode. The performance of this configuration can be seen in Fig. 12-II(a). Figure 12-I(b) shows a configuration very similar to that of Fig. 12-I(a), but using two quantum repeaters, where the second one is highlighted in blue. The outcomes of this configuration can be seen in Fig. 12-II(b). Figure 12-I(c) presents a configuration very similar to the previous one although using a single source generating two disjoint sets of entangled pairs based on a 2-qubit QFT block. The outcomes of this configuration can be seen in Fig. 12-II(c). Finally, Fig. 12-I(d) shows a configuration similar to the two previous cases, but using a single source of the $|GHZ_4\rangle$-type of entangled particles. Figure 12-II(d) shows us the corresponding outcomes of this configuration.

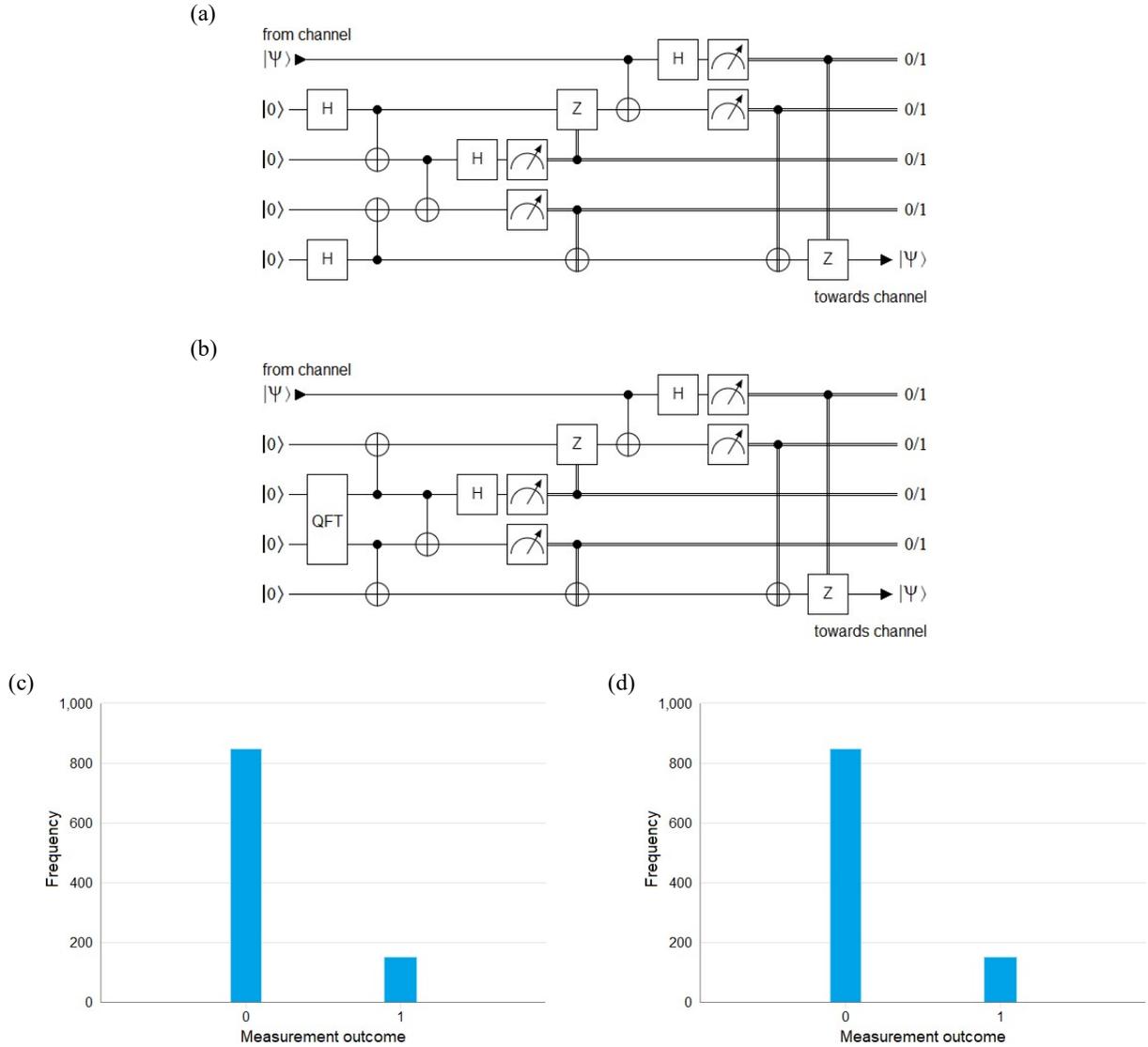

**FIGURE 11** Quantum repeaters via entanglement swapping: a) original configuration based on two sources of $|\beta_{00}\rangle$ entangled states, b) new configuration based on a single source of two independent pairs of $|\beta_{00}\rangle$ entangled states, c) outcomes of the configuration (a) for the state $|\psi_1\rangle$, and d) outcomes of the configuration (b) for the state $|\psi_1\rangle$.

In principle, it is a bit contradictory to talk about extending the range of transmission when the source generating entangled particles is only one and therefore is within the range of the entire chain of quantum repeaters. The question is: In what kind of architecture do these configurations make sense? We will get to this when we analyze the example in subsection "*3.6 An example of multi-orbit architecture*", which involves Figs. 13 and 14 that will allow us to understand the essential need for a source of entangled photons equidistant from a long chain of quantum repeaters of these characteristics.

Table 6 shows the fidelities corresponding to the four experiments in Fig. 12-I. There is a decrease in the performance of the configurations, particularly the last three (b, c, and d) with two quantum repeaters compared to the first (a), which uses only one. This is a direct consequence of the increase in the number of gates. The last three configurations present similar fidelities, although the one that uses the entanglement parallelization technology presents some improvement compared to its closest competitors, being the last one, i.e., the one that uses the $|GHZ_4\rangle$ state, which is the one with the lowest performance.

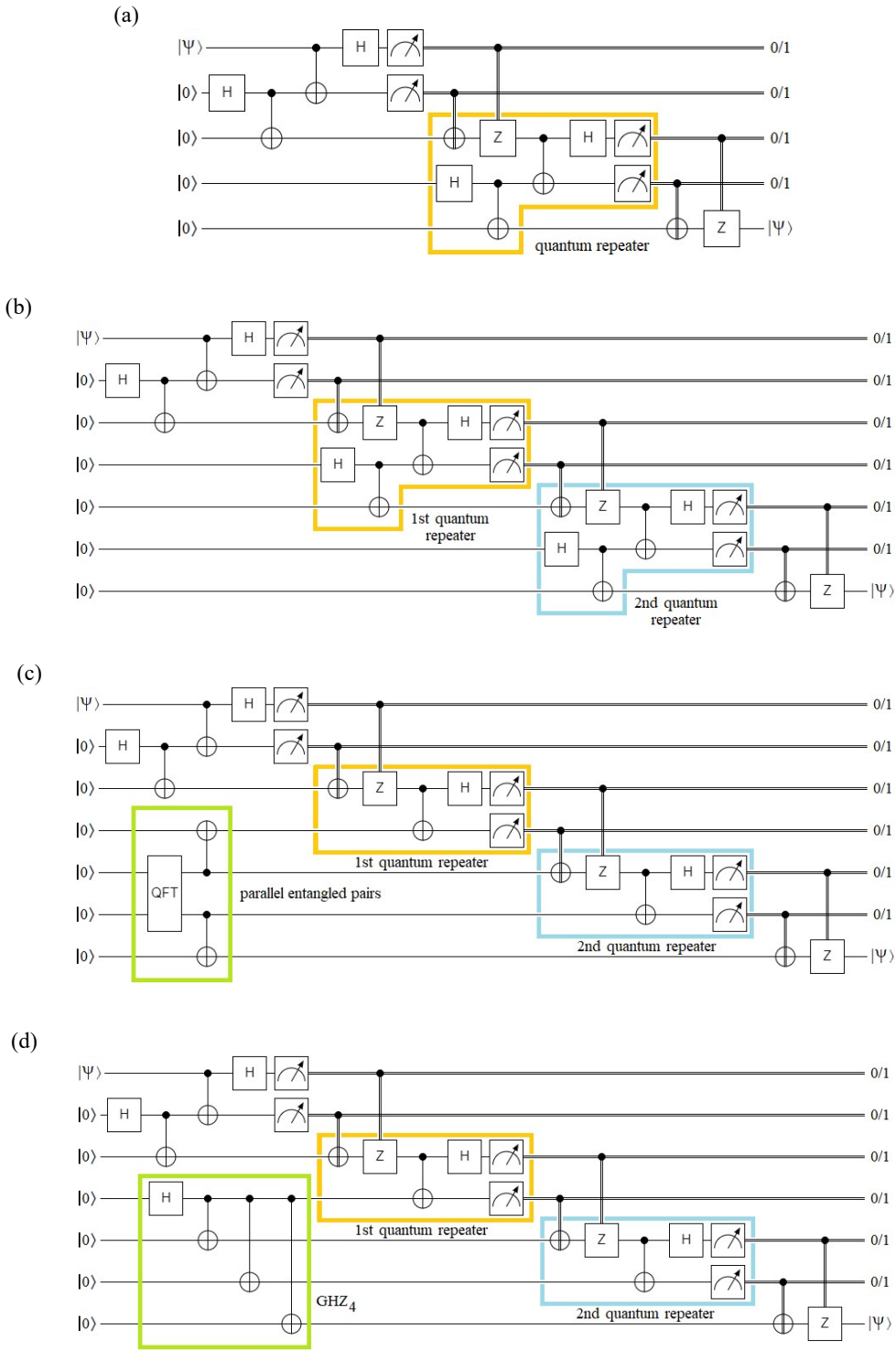

**FIGURE 12-I** Quantum repeaters via the forward cascade technique: a) only one quantum repeater (teleportation highlighted in orange) between sender and receiver, b) two independent quantum repeaters (highlighted in orange and blue), c) two quantum repeaters (highlighted in orange and blue) and a single source (green) of two independent pairs of $|\beta_{00}\rangle$ states, d) two quantum repeaters (highlighted in orange and blue) and a single source (green) of four entangled states of $|GHZ_4\rangle$-type.

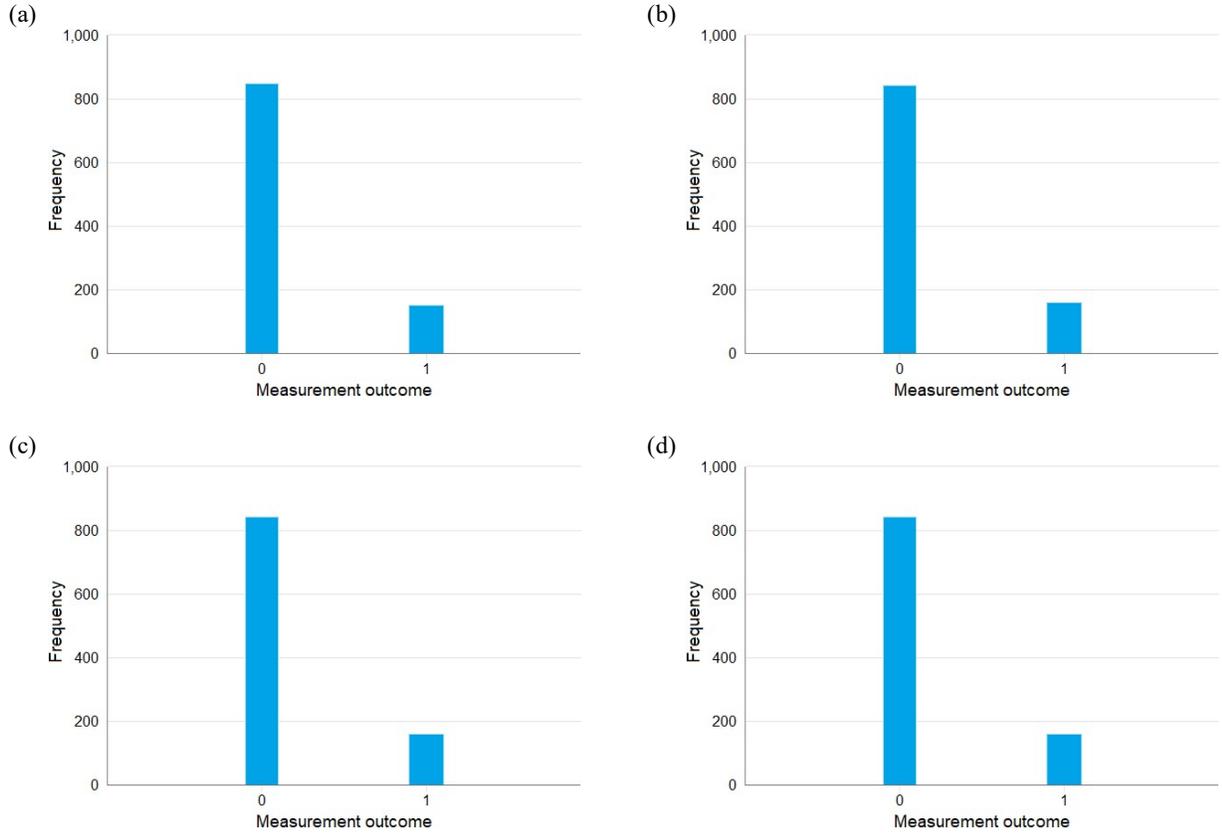

**FIGURE 12-II** Outcomes of the transmitted state $|\psi_1\rangle$ for the four configurations of Fig. 12-I: a) for Fig. 12-I(a), b) for Fig. 12-I(b), c) for Fig. 12_I(c), and d) for Fig. 12-I(d).

**TABLE 6** Fidelities associated with the forward cascade protocols when they transmit the state $|\psi_1\rangle$.

| Version | $F_{RP}$ | $F_{TP}$ | $F_{AP}$ |
|---|---|---|---|
| Figs. 12-I(a), and 12-II(a) | 0.8512 | 0.8409 | 0.8621 |
| Figs. 12-I(b), and 12-II(b) | 0.8407 | 0.8398 | 0.8405 |
| Figs. 12-I(c), and 12-II(c) | 0.8416 | 0.8395 | 0.8411 |
| Figs. 12-I(d), and 12-II(d) | 0.8392 | 0.8201 | 0.8193 |

### 3.5 Comparatively discussing the results

Table 7 represents the compilation of the three fidelities for all the protocols seen so far. The reason for the performance degradation of the different protocols, from those at the top of the table to those at the bottom of the table, has to do with several factors:

a)  The age of the processor and hence its incremental decoherence as the age increases,
b)  the number of gates used by the protocol,
c)  the number of qubits used by the protocol,
d)  the fact that the qubit results from an intermediate instance, i.e., from a reconstruction of a previous transmission, and therefore is not the qubit prepared at the beginning of the protocol, as in the case of the experiments in Fig. 12-I, and

**TABLE 7** Fidelities in terms of the protocol (and its number of qubits), processors (and its number of qubits), and the qubit to be transmitted.

| Protocol | Processor | $|\psi_1\rangle$ | | | $|\psi_2\rangle$ | | | $|\psi_3\rangle$ | | | $|\psi_4\rangle$ | | |
|---|---|---|---|---|---|---|---|---|---|---|---|---|---|
| | | $F_{RP}$ | $F_{TP}$ | $F_{AP}$ | $F_{RP}$ | $F_{TP}$ | $F_{AP}$ | $F_{RP}$ | $F_{TP}$ | $F_{AP}$ | $F_{RP}$ | $F_{TP}$ | $F_{AP}$ |
| Tele 3 qubits | Belem 5 qubits | 0.8782 | 0.8519 | 0.8737 | - | - | - | 0.9865 | 0.9913 | 0.9778 | - | - | - |
| 2-Tele 6 qubits | Oslo 7 qubits | 0.8664 | 0.9522 | 0.9142 | - | - | - | 0.9562 | 0.9416 | 0.9646 | - | - | - |
| 4-Tele 12 qubits | Melbourne 14 qubits | 0.8667 | 0.9264 | 0.8903 | 0.8612 | 0.9298 | 0.9114 | 0.9587 | 0.9415 | 0.9228 | 0.9018 | 0.9125 | 0.9043 |
| QSS 4 qubits | Belem 5 qubits | 0.8212 | 0.9280 | 0.8932 | - | - | - | 0.9712 | 0.9401 | 0.9689 | - | - | - |
| 2-QSS 8 qubits | Melbourne 14 qubits | 0.8209 | 0.9239 | 0.8914 | - | - | - | 0.9641 | 0.9273 | 0.9632 | - | - | - |
| Bidirec 6 qubits | Oslo 7 qubits | 0.8643 | 0.9469 | 0.9174 | - | - | - | 0.9443 | 0.9465 | 0.9678 | - | - | - |
| En-sw 1 5 qubits | Belem 5 qubits | 0.8535 | 0.8478 | 0.8413 | - | - | - | - | - | - | - | - | - |
| En-sw 2 5 qubits | Belem 5 qubits | 0.8562 | 0.8474 | 0.8536 | - | - | - | - | - | - | - | - | - |
| Fo-ca 1 5 qubits | Oslo 7 qubits | 0.8512 | 0.8409 | 0.8621 | - | - | - | - | - | - | - | - | - |
| Fo-ca 2 7 qubits | Oslo 7 qubits | 0.8407 | 0.8398 | 0.8405 | - | - | - | - | - | - | - | - | - |
| Fo-ca 3 7 qubits | Oslo 7 qubits | 0.8416 | 0.8395 | 0.8411 | - | - | - | - | - | - | - | - | - |
| Fo-ca 4 7 qubits | Oslo 7 qubits | 0.8392 | 0.8201 | 0.8193 | - | - | - | - | - | - | - | - | - |

**Where:** Tele = teleportation, 2-Tele = double Tele, 4-Tele = four tele, QSS = quantum secret sharing, 2-QSS = double QSS, Bidirec = bidirectional Tele, En-sw 1 = entanglement swapping, En-sw 2 = entanglement swapping implemented via QFT, and Fo-ca $n$ ($\forall\, n \in [1,4]$) = forward cascade with $n$ quantum repeaters.

e) the type of qubit to be transmitted, i.e., the obvious advantages of transmitting a known and simple qubit like $|0\rangle$, $|1\rangle$, $|+\rangle$, $|-\rangle$, over a generic one as $|\psi_1\rangle$ in terms of performance improvement of the first over the second.

In all the experiments carried out so far, we have used a reduced version of all the protocols (including the example in the next subsection) [71,72], i.e., without quantum measurement modules, since the physical IBM Q machines do not support the *if-then-else* statement, nor the measurement modules in intermediate instances of quantum circuits [43]. For example, the simplified version of the quantum teleportation protocol used in Subsection 3.1 is that of Fig. 13 and not that of Fig. 5(a). Figures 5-12 are only supported by simulators, e.g., Qiskit [43]. The outcomes obtained, as well as the performances related to the fidelities, do not change at all between the original version of Fig. 5(a) and the simplified version of Fig. 13.

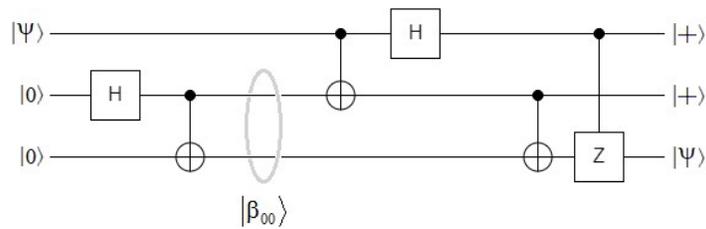

**FIGURE 13** A simplified version of the quantum teleportation protocol supported by all physical IBM Q machines [43].

## 3.6 An example of multi-orbit architecture

Currently, there is a very diverse number of applications that use combinations of satellites and optical ground station (OGS) [73], satellites and drones [74], and satellites with satellites [75], in all cases, exploiting the virtues of quantum entanglement [76]. All these architectures are composed of the so-called Cubesats [77] or more complex satellites. The example presented here is not the exception. Figure 14 presents a multi-orbit architecture that exclusively employs Cubesats [77], where a CubeSat in geostationary orbit (GEO) generates and distributes four independent pairs of entangled photons (quantum channels) thanks to a 4-qubit QFT block. The yellow lines represent the distribution of the mentioned photons between five Cubesats (in reality, and given the dimensions involved, more than five Cubesats are needed to perform this task) in Low Earth Orbit (LEO), that is, approximately 500 km around the Earth, equidistant from each other, in such a way that a communication link can be generated between the points A and B, which are at the planet's antipodes. The orange lines between point A and the first satellite on the left, as well as between satellites, and between the last satellite on the right and point B represent the classic channels, which are required in all teleportation protocols to finish characterizing the teleported state at the destination, since this example uses the quantum repeaters technique of the forward cascade type. In addition, in a quantum key distribution (QKD) context [32-38], entanglement swapping type quantum repeaters expose the key as it passes through the same quantum repeater, which is why the technology selected for this experiment is the best, not only for communication performance, but also from a security point of view. Regarding quantum channels (yellow), the pairs of distributed entangled photons are of the $|\beta_{00}\rangle$-type, while the classical channels (orange) correspond to radio transmissions of the electromagnetic type in *X* or *S* Bands.

In this experiment, we will transmit digital (binary) information, from point A on the planet to point B on the other side of the world, through a multi-orbit architecture, which uses space quantum repeaters based on the Cubesat standard [77]. Therefore, if each Cubesat carries onboard part of a quantum teleportation protocol and due to the simplicity of the states to be teleported, that is, computational basis state (CBS) $\{|0\rangle, |1\rangle\}$, which constitute the quantum version of the classical binary states {0, 1}, then we will turn to an even more simplified version of the quantum teleportation protocol implemented in every Cubesat. This version can be seen in the first four blocks on the right, which range from red to light pink in Fig. 15-I. In these blocks, it can be seen that the module known as BSM has a single quantum measurement block and no Hadamard (H) matrix. This reduced version is more than justifiable, since with the original version the transmission performance drops by half as two classic bits of disambiguation are required to be able to teleport a single CBS, while in the version of the blocks of Fig. 15-I, the relationship is one to one. Figure 15-I shows the complete architecture of this experiment. In the upper-left corner of this figure, specifically at point A (on Earth), the digital (binary) message to be sent in bits is generated, which is transmitted via radio, in *X* or *S* Band, to the first Cubesat (red), where its first sub-block is a classic-to-quantum (Cl2Qu) interface, which allows going from the binary base to its CBS counterpart, i.e., $\{0,1\} \rightarrow [Cl2Qu] \rightarrow \{|0\rangle, |1\rangle\}$.

The color gradient between quantum repeaters, which goes from red in the first Cubesat to very light pink in the last one, has to do with the $F_{AP}$ degradation as the transmitted message progresses through the Cubesat sequence. With each teleportation, the signal degrades, because the next teleportation does not have the original qubit to be teleported, but a previously reconstructed one. As we can see from Fig. 15-I, the latest Cubesat only has a unitary transform consisting of a Pauli matrix *X* [5] and a quantum-to-classic (Qu2Cl) interface, which allows the message to be recovered in binary form from its counterpart on CBS, i.e., $\{|0\rangle, |1\rangle\} \rightarrow [Qu2Cl] \rightarrow \{0,1\}$. Moreover, it is precisely this last satellite that makes the radio link in Band *S* or *X* with the final recipient on Earth (point B). The four independent pairs of entangled photons of the $|\beta_{00}\rangle$-type are generated thanks to a 4-qubit QFT block highlighted in green on board the geostationary satellite. As we can see in Fig. 15-I, the simplified BSM modules do not have quantum measurements that imply the use of the *if-then-else* statement, to be able to implement this experiment on an IBM Q physical machine, in this case, we are talking about the 14-qubit Melbourne processor. Finally, Figs. 15-II(a) and (b) represent the theoretical CBS $\{|0\rangle, |1\rangle\}$ to be transmitted between A and B, respectively, while Figs. 15-II(c) and (d) correspond to the CBS recovered in the last Cubesat, in their respective order. Table 8 shows the $F_{AP}$ fidelity degradation when broadcasting CBS $\{|0\rangle, |1\rangle\}$ via each Cubesat, in such a way that as we move through the Cubesats, the fidelity degrades noticeably from left to right.

This particular experiment constitutes a technical milestone, since it is the first of its kind in which successive quantum repeaters are fed with the same source of entangled photons and that source is at the same distance from each of the respective quantum repeaters. Therefore, the original quantum repeater model based exclusively on entanglement swapping [39, 40, 63-69] has been set aside to replace it with a different technique, simpler and more robust, with fewer gates and qubits and so with greater fidelity when it comes to extending the range of the secure transmission of the CBS packet that constitutes the message, since it is not exposed in the repeater as it happens in entanglement swapping.

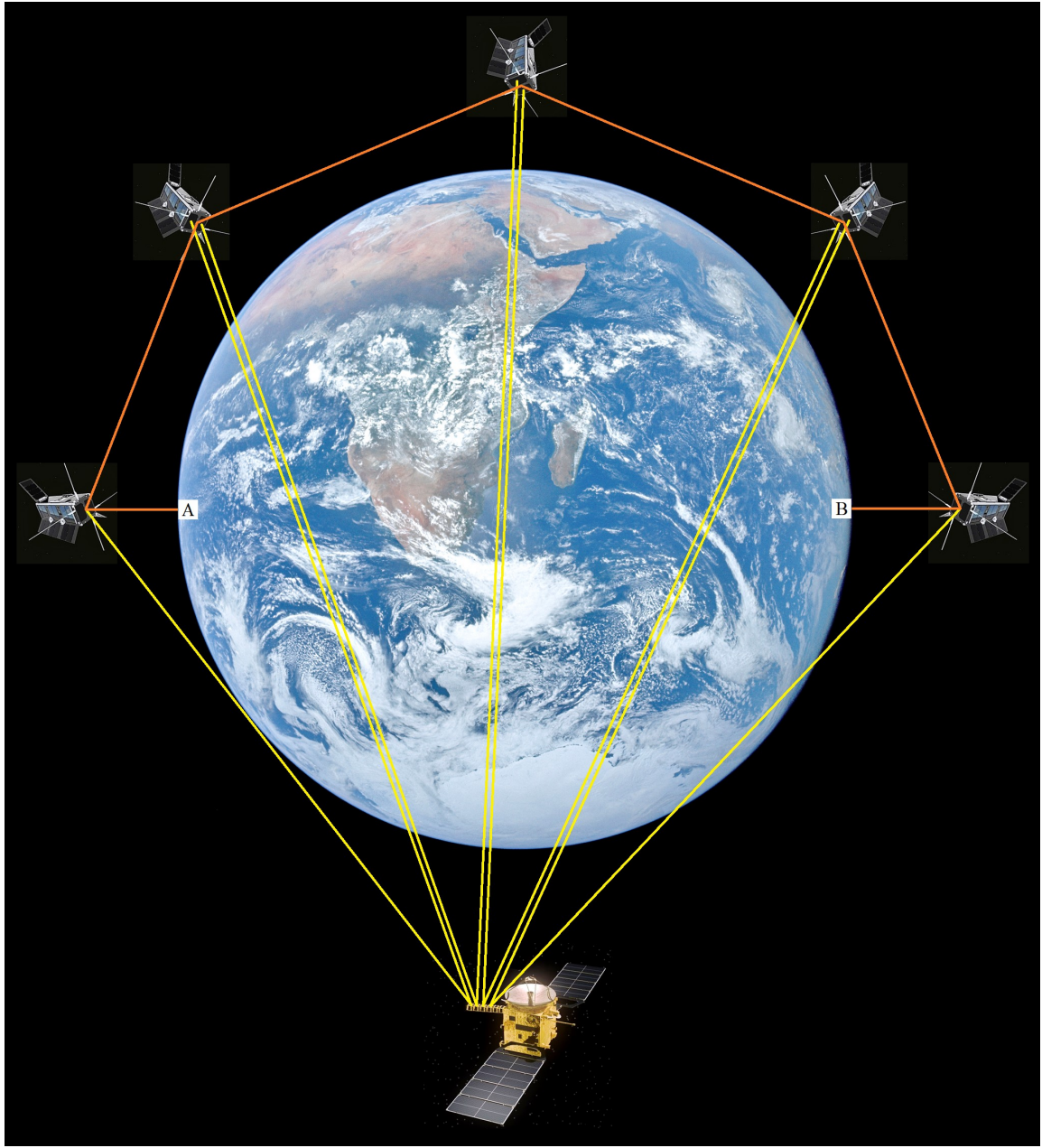

**FIGURE 14** Multi-orbit architecture composed of Cubesats in low Earth orbit (LEO), polar type, and a geostationary satellite. More than five Cubesats are necessary between terrestrial points A and B. The two Cubesats at the ends are responsible for the link with Earth, while the three (more actually) in the middle act as space quantum repeaters. The geostationary satellite contains a single source, which generates four independent pairs of entangled photons of $|\beta_{00}\rangle$-type. Lines in red represent classic channels (generally, radio transmission of Band $X$ or $S$), while yellow lines represent quantum channels, i.e., a beam of photons. Here, we use only three Cubesats in the middle so as not to complicate the implementation on the IBM Q processor.

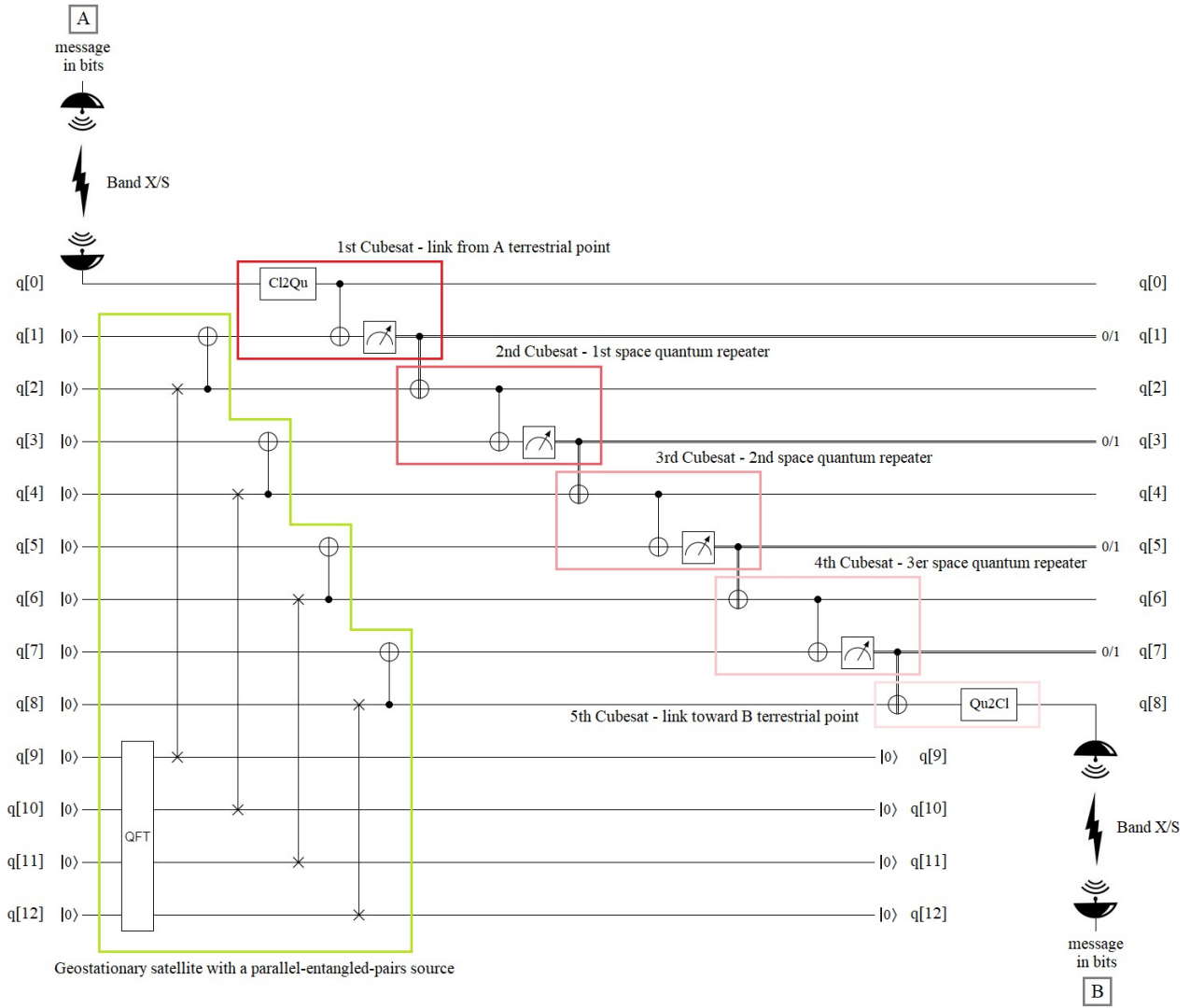

**FIGURE 15-I** Circuital configuration of Fig. 13. A single source of four independent $|\beta_{00}\rangle$ pairs based on a 4-qubit QFT block is on board the geostationary satellite (highlighted in green). Five Cubesats (in a polar orbit, LEO-type) carry out liaison tasks with Earth (uplink, the first one, and downlink, the last one) and space quantum repeaters (the middle three). The colors of the five Cubesats were particularly chosen to represent, from left to right, the degradation of teleportation fidelity of the $|\psi\rangle$ state as it moves through the Cubesats, which turns out to be a computational basis state $\{|0\rangle, |1\rangle\}$. On Earth, we only work with classical (binary) states, for this reason, the first satellite has on board a classic-to-quantum interface (Cl2QU), while the last one contains its counterpart, i.e., a quantum-to-classic interface (Qu2Cl).

## 4 CONCLUSIONS

In this study, a technique for the creation of unique sources generating independent sets of entangled particles based on QFT has been presented. The possibility of having, for the first time in the literature, a technique for the parallelization of entanglement will make it possible to carry out multi-orbit architectures such as those presented in Fig. 15-I, thus extending the scope of the future network of networks, i.e., the Quantum Internet, to all corners of the planet and beyond, thanks to the intervention of quantum repeaters working in the forward cascade mode.

(a) (b)

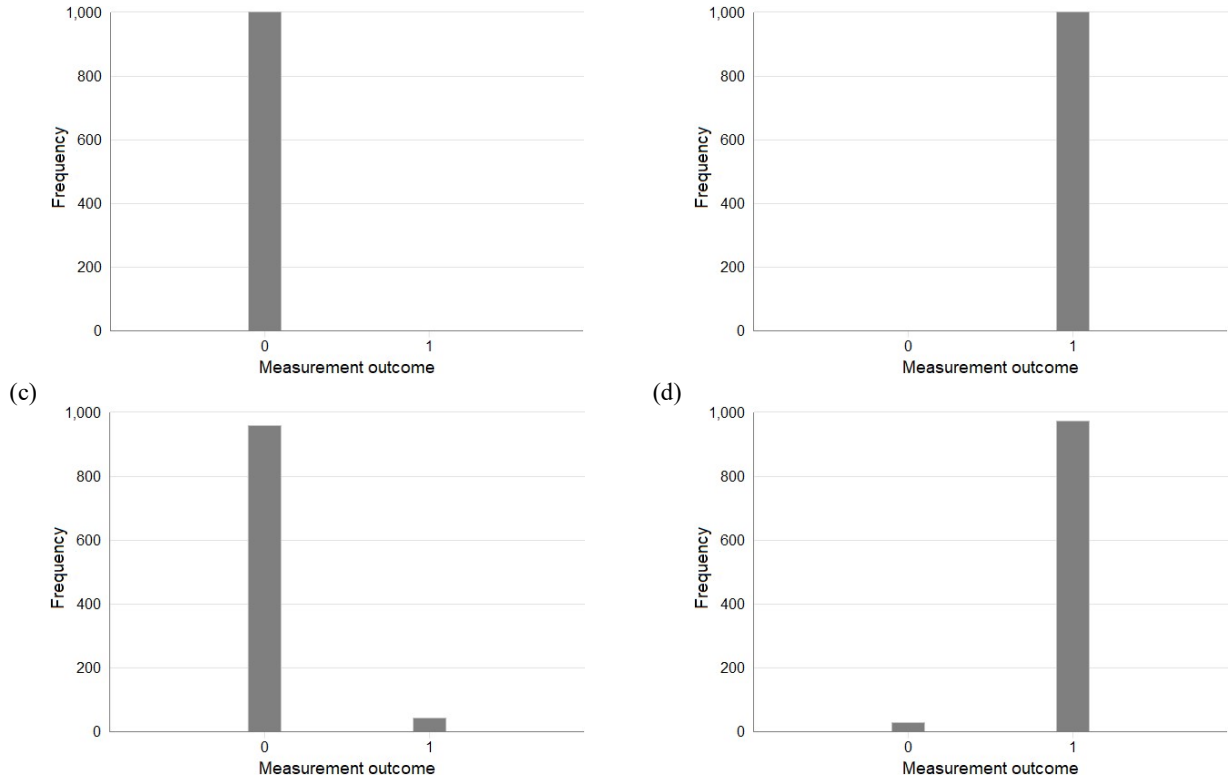

**FIGURE 15-II** Outcomes before and after the transmission of the computational basis states $\{|0\rangle, |1\rangle\}$ for the configuration of Fig. 14-I implemented in a 14-qubit (Melbourne) IBM Q physical platform: a) $|0\rangle$ to be teleported from point A, b) $|1\rangle$ to be teleported from point A, c) outcome obtained at point B when a $|0\rangle$ is transmitted, and d) outcome obtained at point B when a $|1\rangle$ is transmitted. The degradation of fidelity is the result of several factors: decoherence of the platform chosen for the implementation, and inherent errors of the circuit such as bit-flip, phase-flip, and bit-phase-flip.

**TABLE 8** $F_{AP}$ fidelity degradation when broadcasting CBS $\{|0\rangle, |1\rangle\}$ via each Cubesat.

| qubit | after Cubesat #1 | after Cubesat #2 | after Cubesat #3 | after Cubesat #4 | after Cubesat #5 |
|---|---|---|---|---|---|
| $|0\rangle$ | 0.9716 | 0.9524 | 0.9338 | 0.9214 | 0.9078 |
| $|1\rangle$ | 0.9728 | 0.9435 | 0.9327 | 0.9202 | 0.9115 |

This study is completed with numerous application examples of the technique presented with implementations carried out in three physical processors of the IBM Q family [43]. The results obtained show expected and perfectly justifiable results considering different aspects related to the platform on which each experiment was implemented, as well as the most outstanding characteristics of each protocol, that is, its number of gates and qubits involved. Another relevant aspect is the type of qubit to be transmitted (known or generic) and the chaining of successive blocks, which constitutes a determining factor in the degradation of the transmission fidelities involved in each case.

**ACNOWLEDGMENTS**

The author thanks the staff of the Knight Foundation School of Computing and Information Sciences at Florida International University for all their help and support.

**CONFLICT OF INTEREST**

The author declares that he has no competing interests.

**AUTHOR CONTRIBUTIONS**

M.M. conceived the idea, fully developed the theory and experiments, wrote the complete manuscript, prepared figures, and reviewed the manuscript.

**DATA AVAILABILITY STATEMENT**

The experimental data that support the findings of this study are available in ResearchGate with the identifier
https://doi.org/10.13140/RG.2.2.28368.38404